\documentclass[twoside, 11pt, journal, onecolumn]{IEEEtran}

\usepackage{amssymb}
\usepackage{amsmath}
\usepackage{amsthm}
\usepackage[dvips]{graphicx}
\usepackage{subfig}
\usepackage{url}
\usepackage{setspace}

\theoremstyle{plain}
\newtheorem{theorem}{Theorem}
\newtheorem{lemma}{Lemma}
\newtheorem{proposition}{Proposition}
\newtheorem{corollary}{Corollary}
\newtheorem{conjecture}{Conjecture}

\theoremstyle{definition}
\newtheorem{definition}{Definition}

\newtheorem{remark}{Remark}

\begin{document}

\markboth{SUBMITTED TO THE IEEE TRANSACTIONS ON INFORMATION THEORY,
MAY 2010.}{E. HOF and S. Shamai:
Secrecy-Achieving Polar-Coding for Binary-Input Memoryless Symmetric Wire-Tap Channels}

\title{\Huge Secrecy-Achieving Polar-Coding for Binary-Input Memoryless Symmetric Wire-Tap Channels}

\author{\vspace{1cm} \IEEEauthorblockN{Eran Hof\footnote{Eran Hof is the corresponding author (E-mail: eran.hof@gmail.com).} \ \ \ Shlomo
Shamai}\\[0.2cm]
\IEEEauthorblockA{Department of Electrical Engineering\\
Technion -- Israel Institute of Technology\\
Haifa 32000, Israel \\
}}

\date{}
\maketitle

%\hspace{7.4cm} \today \vspace*{0.5cm}

\begin{abstract}
A polar coding scheme is introduced in this paper for the wire-tap
channel. It is shown that the provided scheme achieves the entire rate-equivocation region for the case of symmetric and degraded wire-tap channel, where the weak notion of secrecy is assumed. For the particular case of binary erasure wire-tap channel, an alternative proof is given. The case of general non-degraded wire-tap channels is also considered.
\end{abstract}

\IEEEpeerreviewmaketitle

\section{Introduction}

Channel coding via the method of channel polarization is provided by
Arikan in~\cite{ArikanPolarCodes}. On a binary-input discrete
memoryless channel (DMC), polarization ends up with either `good
bits', i.e., binary channels whose capacity approaches~1 bit per
channel use, or `wasted bits', i.e., channels whose capacity
approaches zero. The fraction of the good bits is equal to the
mutual information with equiprobable inputs (which equals the
capacity for the case of symmetric channels). In a physically
degraded setting, as mentioned in~\cite{ArikanPolarCodes}, an order
of polarization is maintained in the sense that `good' bits for the
degraded channel, must also be `good' for the better channel.

For a standard single-user channel coding problem, the polar coding
scheme is based on transmitting the uncoded information bits over
the capacity approaching channels (when we interpret the
polarization as a kind of a precoding or pre-processing). At the
same time, fixed and predetermined bits are transmitted over the
channels whose capacity approaches zero. These predetermined bits
are still needed in the successive decoding process, hence they can
not be ignored.

A secrecy polar scheme is suggested in this paper for the
wire-tap channel. A secret message needs to be transmitted reliably
to a legitimate user. At the same time, this message must be kept
secret from the eavesdropper. At the first part of this paper, it is
assumed that the marginal channel to the eavesdropper is physically
degraded with respect to the marginal channel to the legitimate
user. The proposed secrecy polar scheme for the degraded case is
based on transmitting random bits on the `good bits' of the degraded
eavesdropper channel. These random bits are independent of the
secret message. At the legitimate receiver, the random bits can be
decoded reliably. This is because the `good bits' for the degraded
eavesdropper channel are also `good' for the legitimate user. The
rest of the `good' bits for the legitimate user are dedicated for
the secret message. Additional independent works on this subject are provided in~\cite{theIndependentVardyWork}~\cite{theOtherIndependentVardyWork}~\cite{theThirdIndependentWork}.

Transmitting random bits on the `good bits' of the eavesdropper, all
the possible information rates that can be detected by the
eavesdropper are exhausted. Otherwise, the standard channel capacity
could have been beaten. Thus the `good bits' associated with the
secret message for the legitimate channel, must be perfectly secret
(at least in the weak sense). Note that this result is satisfied
immaterial of whether the eavesdropper adheres to successive
decoding or to optimal decoding (as otherwise, its capacity could
have been beaten). It is first shown that the provided scheme archives the secrecy capacity for the considered model. The result is then generalized to the enitre rate-equivocation region. This result is proved under a weak notion of secrecy. For the particular case of binary erasure wire-tap channel, an alternative proof is provided, based on algebraic arguments. This different notion of proof may contribute to a stronger notion of security.

At the second part of the paper, the secrecy polar scheme is adapted
for the general, i.e., non-degraded wire-tap channel. This scheme is
based on a conjecture on possible polarization properties of some of
the `bad' indices of the eavesdropper. To this end, the polarization
of the 'bad' bits is concerned while the decoder has perfect
knowledge of some of the 'good' bits (which are no longer part of
the transmitted message, but they are kept predetermined and fixed).
The question regarding this aspect is whether the additional
information helps in decoding these bits or do they keep their
original 'bad' polarization. The original polarization result
in~\cite{ArikanPolarCodes} does not fully cover this scenario.
%Two
%possible suggestion for this polarization are conjectured in this
%part of the draft. Assuming that at least one of this possible
%polarization properties is correct, allows for an adaptation of the
%polar secrecy scheme for the non-degraded case.

This paper is structured as follows: In
Section~\ref{section:preliminaries} preliminary introduction is
provided. In Section~\ref{section:communicationModel} the wire-tap
communication model is introduced in addition to some basic
definitions and results in information-theoretic security. Polar
codes are introduced in Section~\ref{section:polarCodes}. The polar
secrecy scheme is detailed and studied in
Section~\ref{section:TheProposedScheme}. A conjecture on possible
polarization properties is stated in
Section~\ref{section:proposedScheme4GeneralWireTapeCh}, along with a
resulting adaptation of the polar secrecy scheme for non-degraded
wiretap channels. A list of possible further generalizations is
provided in Section~\ref{section:generalizations}.

\section{Preliminaries}
\label{section:preliminaries}

\subsection{The Wire-Tap Communication Model}
\label{section:communicationModel}

\begin{figure}[here!]
\centering
\includegraphics[width=5in]{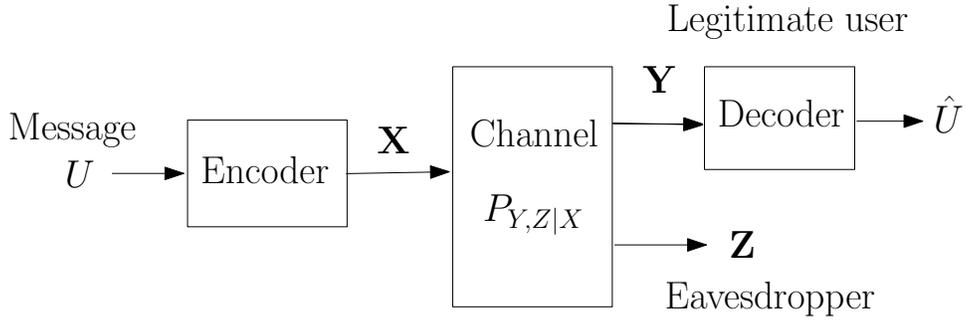}
\vspace*{-1cm}\caption{\footnotesize{A wire-tap communication
model.}}
 \label{figure:wireTapCommunicationModel}
\end{figure}

We consider the communication model in
Figure~\ref{figure:wireTapCommunicationModel}. A coded system is
presented which transmits a confidential message $U$ to a legitimate
user. The message $U$ is chosen uniformly from a set of size $M$.
Next, the message is encoded to a codeword $\mathbf{X}$ with a
blocklength $n$ over an alphabet $\mathcal{X}$. The resulting
code-rate is $R = \frac{1}{n}\log M$. The codeword $\mathbf{X}$ is
transmitted over a communication channel $P_{Y,Z|X}$ with one input,
and two outputs. The transmission is assumed to take place over a
DMC $P$, with an input alphabet $\mathcal{X}$, and output alphabets
$\mathcal{Y}$ and $\mathcal{Z}$. Let
$P(\mathbf{y},\mathbf{z}|\mathbf{x})$ denote the probability of
receiving the vectors $\mathbf{y} \in \mathcal{Y}^n$, and
$\mathbf{z} \in \mathcal{Z}^n$, at the legitimate user and the
eavesdropper, respectively, given that a codeword
$\mathbf{x}\in\mathcal{X}^n$ is transmitted. Based on the assumption
that the channel is memoryless, it follows that
\begin{equation*}
P(\mathbf{y},\mathbf{z}|\mathbf{x}) = \prod_{k=1}^n P(y_k,z_k|x_k)
\end{equation*}
where (with some abuse of notation) $P(y,z|x)$ denotes the
probability of receiving the symbols $y \in \mathcal{Y}$ and
$z\in\mathcal{Z}$, at the legitimate user and the eavesdropper,
respectively, given that the symbol $x \in \mathcal{X}$ is
transmitted. Moreover, let $G(y|x)$ and $Q(z|x)$ denote the marginal
probabilities for receiving the symbols $y \in \mathcal{Y}$ and
$z\in\mathcal{Z}$, at the legitimate user and the eavesdropper,
respectively, given that the symbol $x \in \mathcal{X}$ is
transmitted. Both $G(y|x)$ and $Q(z|x)$ are transition probability
laws of DMCs, called the marginal channels of the legitimate user
and the eavesdropper, respectively. In addition, the probability to
receive the symbol $z\in\mathcal{Z}$ at the eavesdropper, given that
the symbol $y \in \mathcal{Y}$ is received at the legitimate user is
denoted by $D(z|y)$.

The channel output vectors $\mathbf{Y}$ and $\mathbf{Z}$, both of
length $n$, are received by the legitimate user and the
eavesdropper, respectively. The legitimate user decodes the received
vector $\mathbf{Y}$ resulting in the decoded message $\hat{U}$. The
objectives of the considered coding system is to obtain both secure
and reliable communication. These objectives are to be accomplished
simultaneously using a single codebook $\mathcal{C}_n$. The
reliability of the system is measured via the average error
probability $P_{\text{e}}(\mathcal{C}_n)$ of the decoded message
\begin{equation*}
P_{\text{e}}(\mathcal{C}_n) = \frac{1}{M} \sum_{m=1}^M \Pr
\left(\hat{U}\neq m |\ U = m\right).
\end{equation*}
Note that the error probability depends on the blocklength of the
coded message. The level of security is measured by the equivocation
rate
\begin{equation} \label{equation:equivocationRate}
R_{\text{e}}(\mathcal{C}_n) \triangleq \frac{1}{n} H(U|\mathbf{Z})
\end{equation}
where $H(U|\mathbf{Z})$ denotes the conditional entropy of the
transmitted message $U$, given the received vector $\mathbf{Z}$ at
the eavesdropper.

\begin{definition}[\textbf{Achievable rate-equivocation pair}] \label{definition:achievablePair}
A rate-equivocation pair $(R,R_{\text{e}})$ is achievable if there
exists a code sequence $\{\mathcal{C}_n\}$ of block length $n$ and
rate $R$ such that
\begin{align*}
& \lim_{n\to \infty} P_{\text{e}}(\mathcal{C}_n) = 0\\
& R_{\text{e}} \leq \lim_{n\to\infty} R_{\text{e}}(\mathcal{C}_n).
\end{align*}
\end{definition}

\begin{remark}[\textbf{On strong and weak notions of secrecy}]
\label{remark:strongWeekSecrecy} The current discussion considers
normalized entropies to measure the level of security (see the
definition of equivocation rate
in~\eqref{equation:equivocationRate}). Therefore, the achieved
secrecy notion is referred to as {\em weak} secrecy. The {\em
strong} notion of secrecy considers the unnormalized mutual
information between the confidential message and the received vector
at the eavesdropper receiver. Strong secrecy guarantees secrecy in
the weak sense while the opposite direction does not follow.
\end{remark}

\begin{definition}[\textbf{Secrecy capacity}]
The secrecy capacity $C_{\text{s}}$ is the supremum of all the rates
$R$, such that the pair $(R,R)$ is an achievable rate-equivocation
pair.
\end{definition}

\begin{theorem}[\textbf{The secrecy capacity of the wire-tap
channel \cite{InformationTheoreticSecurityTutorial}}] {\em
\label{theorem:generalWireTapCapacity} The secrecy capacity
$C_{\text{s}}$ of the wire-tap channel satisfies:
\begin{equation*}
C_{\text{s}} = \max_{P_{UX}P_{YZ|X}} \bigl( I(U;Y) - I(U;Z) \bigr)
\end{equation*}
where $U$ is an auxilary random variable over the alphabet
$\mathcal{U}$, satisfying
\begin{enumerate}
\item Markov relationship: $U\to X\to (Y,Z)$ is a Markov chain.
\item Bounded cardinality: $|\mathcal{U}| \leq \mathcal{X}+1$.
\end{enumerate}}
\end{theorem}

Binary-input symmetric wire-tap channels are considered in this
paper.

\begin{definition}[\textbf{Symmetric binary input channels}]
\label{definition:symmetricBDMC} A DMC with a transition probability
$p$, binary-input alphabet $\mathcal{X}$, and an output alphabet
$\mathcal{Y}$ is said to be symmetric if there exists a permutation
$\pi$ over $\mathcal{Y}$ such that
\begin{enumerate}
\item The inverse permutation $\pi^{-1}$ is equal to $\pi$, i.e.,
\begin{equation*}
\pi^{-1}(y) = \pi(y)
\end{equation*}
for all $y\in\mathcal{Y}$.
\item The transition probability $p$ satisfies
\begin{equation*}
p(y|0) = p(\pi(y)|1)
\end{equation*}
for all $y\in\mathcal{Y}$.
\end{enumerate}
\end{definition}

%\begin{remark}
%The capacity of a symmetric, binary-input DMC is attained with an
%equiprobability input.
%\end{remark}

\begin{definition}[\textbf{Symmetric binary-input wire-tap
channels}] \label{definition:symmetricBDMWireTapChannel} A binary
input discrete memoryless wire-tap channel is symmetric if both of
its marginal channels are symmetric.
\end{definition}

The particular case of physically degraded channels is studied in
this paper.

\begin{definition}[\textbf{Physically degraded channels}]
Let $P$ be a wire-tap channel with an input alphabet $\mathcal{X}$
and output alphabets $\mathcal{Y}$ and $\mathcal{Z}$, at the
legitimate and eavesdropper, respectively. Then, $P$ is said to be
physically degraded if
\begin{equation} \label{equation:degradationProperty}
P(y,z|x) = G(y|x) D(z|y)
\end{equation}
for all $x\in\mathcal{X}$, $y\in\mathcal{Y}$, and $z\in
\mathcal{Z}$.
\end{definition}

The following Theorem characterizes the secrecy capacity of a
binary-input, memoryless, symmetric and degraded wire-tap channel:

\begin{theorem}[\cite{InformationTheoreticSecurityTutorial}] \label{theorem:secrecyCapacity}{\em
Let $P$ be a binary-input, memoryless, symmetric, and degraded
wire-tap channel. Denote by $G_{Y|X}$ and $Q_{Z|X}$ the marginal
channels to the legitimate user and the eavesdropper, respectively.
Then, the secrecy capacity $C_{\text{s}}$ is given by
\begin{equation*}
C_{\text{s}}(P) = C(G_{Y|X}) - C(Q_{Z|X})
\end{equation*}
where $C(G_{Y|X})$ and $C(Q_{Z|X})$ are the channel capacities of
the marginal channel $G_{Y|X}$ and $Q_{Z|X}$, respectively.}
\end{theorem}

\begin{remark}[\textbf{On the entire rate-equivocation region}]
Theorem~\ref{theorem:secrecyCapacity} is a particular case of the rate-equivocation region of less-noisy channels (which is on its own a particular case of the rate-equivocation region of the wire-tap channel). Under the notation in Theorem~\ref{theorem:generalWireTapCapacity}, if $I(U;Y) \geq I(U;Z)$ for every $U$ satisfying the Markov relationship in Theorem~\ref{theorem:generalWireTapCapacity}, then the channel to the legitimate receiver is said to be {\em less noisy} than the eavesdropper (the degradation assumption in~\eqref{equation:degradationProperty} satisfies the less noisy condition). It can be shown for the case of less-noisy wire-tap channels, that the rate-equivocation region is given by
\begin{equation*}
\bigcup_{P_XP_{YZ|X}} \left\{ (R,R_{\text{e}}):\ \begin{array}{l} 0 \leq R \leq I(X;Y) \\ 0 \leq R_{\text{e}} \leq R \\ R_{\text{e}} \leq I(X;Y) - I(X;Z) \end{array}\right\}.
\end{equation*}
For further details and proof see~\cite{InformationTheoreticSecurityTutorial} and references therein. In the particular case of binary-input, memoryless symmetric and degraded wire-tap channels as in Theorem~\ref{theorem:secrecyCapacity}, the rate-equivocation region is therefore given by
\begin{equation}\label{equation:rateEquivocationRegion}
\left\{ (R,R_{\text{e}}):\ \begin{array}{l} 0 \leq R \leq C(G_{Y|X}) \\ 0 \leq R_{\text{e}} \leq R \\ R_{\text{e}} \leq C(G_{Y|X}) - C(Q_{Z|X}) \end{array}\right\}.
\end{equation}
\end{remark}

\subsection{Polar Codes}
\label{section:polarCodes}

This preliminary section offers a minimal summary of the basic
definitions and results in \cite{ArikanPolarCodes},
\cite{TelatarArikanPolarRate}, that are essential to the
presentation of the results in
Section~\ref{section:TheProposedScheme}.

Let $p$ be a transition probability function of a DMC with a binary
input-alphabet $\mathcal{X} = \{0,1\}$ and an output alphabet
$\mathcal{Y}$. The operation of the channel on vectors is also
denoted by $p$, that is for $\mathbf{x} = (x_1,\ldots,x_n) \in
\mathcal{X}^n$, and $\mathbf{y} = (y_1,\ldots,y_n) \in
\mathcal{Y}^n$, the block transition probability is given by
\begin{equation*}
p(\mathbf{y}|\mathbf{x}) = \prod_{l=1}^n p(y_l|x_l).
\end{equation*}

Polar codes are defined in this section using the following
recursive construction. At the first step, two independent copies of
$p$ are combined to form a new channel $p_2$ over an input alphabet
$\mathcal{X}^2$ and output alphabet $\mathcal{Y}^2$. The transition
probability function of the combined channel is given by
\begin{equation}
\label{equation:FirstStepRecursiveConstructionOfPolarCodes}
p_2(y_1,y_2|u_1,u_2) = p(y_1|w_1 + w_2) p(y_2|w_2)
\end{equation}
for all $y_1, y_2 \in \mathcal{Y}$, and $w_1,w_2 \in \mathcal{X}$,
where the addition operation is carried modulo~2. At the $i$-th step
of the construction, the transition probability function $p_n$, $n=
2^i$, is defined for a combined channel with an input alphabet
$\mathcal{X}^n$ and an output alphabet $\mathcal{Y}^n$. The
recursive definition of $p_n$ is based on two independent copies of
the channel $p_{\frac{n}{2}}$ defined at the previous step ($i-1$).
The channel $p_{\frac{n}{2}}$ has an input alphabet
$\mathcal{X}^{\frac{n}{2}}$ and an output alphabet
$\mathcal{Y}^{\frac{n}{2}}$. The construction of the channel $p_n$
includes the following steps:
\begin{enumerate}
\item
\label{step:ConstructingTheSVectorInPolarizationCodes} An input
vector $\mathbf{w}= (w_1,\ldots,w_n) \in \mathcal{X}^n$ is first
transformed to a vector $\mathbf{s} = (s_1,\ldots,s_n) \in
\mathcal{X}^n$ where
\begin{equation*}
s_{2k-1} = w_{2k-1} + w_{2k}
\end{equation*}
and
\begin{equation*}
s_{2k} = x_{2k},\ \ 1 \leq k \leq \frac{n}{2}
\end{equation*}
where the addition is carried modulo 2.
\item
\label{step:ConstructingTheVVectorInPolarizationCodes} The vector
$\mathbf{s}$ is transformed into a vector $\mathbf{v} \in
\mathcal{X}^n$ where
\begin{equation*}
\mathbf{v} = (s_1,s_3,\ldots,s_{n-1},s_2,s_4,\ldots,s_{n}).
\end{equation*}
i.e., the first $\frac{n}{2}$ elements of $\mathbf{v}$,
$v_1,\ldots,v_{\frac{n}{2}}$ equal the elements in $\mathbf{s}$ with
odd indices, and the rest $\frac{n}{2}$ elements of $\mathbf{v}$,
$v_{\frac{n}{2}+1},\ldots, v_{n}$ equal the elements of $\mathbf{s}$
with even indices. This operation is called a reverse shuffle
operation and can be described by the linear transformation
\begin{equation*}
\mathbf{v} = \mathbf{s} R_n
\end{equation*}
where $R_n$ is an $n \times n$ matrix, called the reverse shuffle
operator.
\item $p_n(\mathbf{y}|\mathbf{w})$ is given by
\begin{equation} \label{equation:combinedChannel}
p_n(\mathbf{y}|\mathbf{w}) = p_{\frac{n}{2}}
\left(\left(y_1,y_2,\ldots,y_{\frac{n}{2}}\right)|\left(v_1,v_2,\ldots,v_{\frac{n}{2}}\right)\right)
p_{\frac{n}{2}}
\left(\left(y_{\frac{n}{2}+1},y_{\frac{n}{2}+2},\ldots,y_{n}\right)|\left(v_{\frac{n}{2}+1},v_{\frac{n}{2}+2},\ldots,v_{n}\right)\right).
\end{equation}
\end{enumerate}
The recursive channel-synthesizing operation of $p_n$ is referred to
as channel combining, and the channel $p_n$ is referred to as the
combined channel. Note that the resulting block length $n$ for this
construction must be a power of $2$, that is $n=2^i$ for
$i\in\mathbb{N}$. Throughout this paper, all block lengths $n$ are
assumed to be integral powers of~2.

The recursive construction of $p_n$ can be equivalently defined
using a linear encoding operation. Let
\begin{equation*}
F = \left( \begin{array}{cc} 1 & 0 \\ 1 & 1 \end{array}\right)
\end{equation*}
and define the following recursive construction of the $n\times n$
matrices $G_n$:
\begin{align}
G_1 = & I_1 \nonumber\\
\label{equation:GRecursiveDefinition} G_n = & \left(I_{\frac{n}{2}}
\otimes F\right) R_n \left(I_2 \otimes G_{\frac{n}{2}}\right)
\end{align}
where $I_l$ is the $l\times l$ identity matrix and $\otimes$ denotes
the Kronecker product for matrices.  The matrix $G_n$ is refereed to
as the polar generator matrix of size $n$.

\begin{proposition}[\cite{ArikanPolarCodes}]
{\em Let $p$ be a DMC, and let $p_n$ be the combined channel with a
block length $n$. Then,
\begin{equation} \label{equation:equivalentRecursiveConstruction}
p_n(\mathbf{y}|\mathbf{w}) = p(\mathbf{y}|\mathbf{w}G_n)
\end{equation}
for all $\mathbf{y} \in \mathcal{Y}^n$ and $\mathbf{w}\in
\mathbf{X}^n$, where $p_n$ is the combined channel
in~\eqref{equation:combinedChannel} and $G_n$ is the $n\times n$
matrix defined in \eqref{equation:GRecursiveDefinition}.}
\end{proposition}

Denote by $[n] \triangleq \{1,2,\ldots,n\}$, and let $\mathcal{A}_n
\subseteq [n]$. In addition, denote by $\mathcal{A}_n^{\text{c}}$
the complementary set of $\mathcal{A}_n$, that is
$\mathcal{A}_n^{\text{c}} = [n] \setminus \mathcal{A}_n$. Given a
set $\mathcal{A}_n$, a class of coset codes with a common code-rate
$\frac{1}{n}|\mathcal{A}_n|$ are formed. Over the indices specified
by $\mathcal{A}_n$, the components of the input vector $\mathbf{w}$
are set according to the information bit vector. The rest of the
bits of $\mathbf{w}$ are predetermined and fixed according to the
particular code design. By setting both the set $\mathcal{A}_n$ and
the components of $\mathbf{w}$ specified by
$\mathcal{A}_n^{\text{c}}$, a particular coset code is defined. This
code can be shown to be a block coset code. The set $\mathcal{A}_n$
is referred as the information set. Polar codes are constructed by a
specific choice of the information set $\mathcal{A}_n$. Moreover,
the choice of the information set is tailored to the specific
channel over which the communication takes place.

A coset code is defined using a linear block code and a coset
vector. Let $G$ be a generator matrix for a binary $(n,k)$ linear
block code with block length $n$ and dimension $k$. In addition, let
$\mathbf{c} \in \mathcal{X}^n$ be a binary vector. Then, the coset
block code $\mathbf{C}(G,\mathbf{c})$ is defined by
\begin{equation} \label{equation:generalCosetCode}
\mathcal{C}(G,\mathbf{c}) \triangleq \left\{ \mathbf{x}:\ \mathbf{x}
= \mathbf{u}G+\mathbf{c}, \ \mathbf{u} \in \mathcal{X}^k\right\}.
\end{equation}
Denote by $G_n(\mathcal{A}_n)$ the $|\mathcal{A}_n| \times n$
sub-matrix of $G_n$, defined by the rows of $G_n$ whose indices are
in $\mathcal{A}_n$. Similarly, the matrix
$G_n(\mathcal{A}^{\text{c}})$ denotes the
$|\mathcal{A}_n^{\text{c}}| \times n$ sub-matrix of $G_n$ formed by
the remaining rows of $G_n$. For each choice of $\mathcal{A}_n$ and
an arbitrary $n-k$ binary vector $\mathbf{b} \in \mathbf{X}^{n-k}$,
a coset code $\mathcal{C}$ is defined according to
\begin{equation} \label{equation:cosetCodeForPolarConstruction}
\mathcal{C} = \mathcal{C}\bigl(G_n\left(\mathcal{A}_n\right),
\mathbf{b} G_n \left(\mathcal{A}_n^{\text{c}}\right)\bigr).
\end{equation}
This code coincides with the recursive construction in
\eqref{equation:equivalentRecursiveConstruction}. To see this, plug
the information vector $\mathbf{u}$ in the information indices,
specified by $\mathcal{A}_n$, of the input vector $\mathbf{w}$ to
the recursive construction. In addition, plug the vector
$\mathbf{b}$ in the rest of the components of $\mathbf{w}$ .

Channel splitting is another important operation that is introduced
in~\cite{ArikanPolarCodes} for polar coding. The split channels
$\{p_n^{(l)}\}_{l=1}^n$, with a binary input alphabet $X$ and output
alphabets $\mathcal{Y}^n \times \mathcal{X}^{l-1}$, $1\leq l\leq n$,
are defined according to:
\begin{equation} \label{equation:standradPolarChannelSplitting}
p_n^{(l)}(\mathbf{y},\mathbf{w}|x) =
\frac{1}{2^{n-1}}\sum_{\mathbf{c} \in \mathcal{X}^{n-l}}
p_n\bigl(\mathbf{y}|(\mathbf{w},x,\mathbf{c})\bigr)
\end{equation}
where $\mathbf{y} \in \mathcal{Y}^n$, $\mathbf{w} \in
\mathcal{X}^{l-1}$, and $x \in \mathcal{X}$. The channel
synthesizing operation in
\eqref{equation:standradPolarChannelSplitting} is referred to as
channel splitting operation. The Bhattacharyya parameter of
$p_n^{(l)}$ is denoted by:
\begin{equation} \label{equation:BhattacharyyaParameter}
B(p_n^{(l)}) \triangleq \sum_{\mathbf{y}\in\mathcal{Y}^n}
\sum_{\mathbf{w} \in \mathcal{X}^{l-1}}
\sqrt{p_n^{(l)}(\mathbf{y},\mathbf{w}|0)
p_n^{(k)}(\mathbf{y},\mathbf{w}|1)}.
\end{equation}
The construction of the sequence of sets of split channels
$\{p_n^{(l)}(\mathbf{y},\mathbf{w}|x)\}_{l=1}^n$, $n=2^i$, $i \in
\mathbb{N}$, in \eqref{equation:standradPolarChannelSplitting} can
be described using the following alternative recursion:

\begin{proposition}[\cite{ArikanPolarCodes}]\label{proposition:splitRecursion}
{\em For all $i>0$, $1\leq l \leq 2^i$,
\begin{align}
\label{equation:oddTransformation4Splitchannel}
p_{2^{i+1}}^{(2l-1)}\bigl((\mathbf{y}^{(1)},\mathbf{y}^{(2)}),\mathbf{w}|w_1\bigr)
= & \sum_{w \in \mathcal{X}} \frac{1}{2} p_{2^i}^{(l)}
\bigl(\mathbf{y}^{(1)},g(\mathbf{w})|w_1 + w\bigr) p_{2^i}^{(l)}\bigl(\mathbf{y}^{(2)}, e(\mathbf{w})|w\bigr)\\
\label{equation:evenTransformation4Splitchannel}
p_{2^{i+1}}^{(2l)}\bigl((\mathbf{y}^{(1)},\mathbf{y}^{(2)}),(\mathbf{w},w_1)|w_2\bigr)
= & \frac{1}{2} p_{2^i}^{(l)}
\bigl(\mathbf{y}^{(1)},g(\mathbf{w})|w_1 + w_2\bigr)
p_{2^i}^{(l)}\bigl(\mathbf{y}^{(2)},e(\mathbf{w})|w_2\bigr)
\end{align}
where $\mathbf{y}^{(1)}, \mathbf{y}^{(2)} \in \mathcal{Y}^{2^i}$,
$\mathbf{w} = (w_1, \ldots, w_{2l-2}) \in \mathcal{X}^{2l-2}$, $w_1,
w_2 \in \mathcal{X}$, the addition operation is carried modulo 2 and
$\mathbf{g} = (g_1,\ldots, g_{l-1}) = g(\mathbf{w})$ is a vector in
$\mathcal{X}^{l-1}$ defined according to
\begin{equation} \label{equation:bitWizeOddEvenXorDefinition}
g_j = w_{2j-1} + w_{2j},\ \ 1\leq j \leq {l-1}
\end{equation}
and
\begin{equation} \label{equation:EvenVectorDefinition}
e(\mathbf{w}) = (w_2,w_4,\ldots,w_{2l-2})
\end{equation}
is the vector in $\mathcal{X}^{l-1}$ comprises from the components
of $\mathbf{x}$ with even indices.}
\end{proposition}

The importance of channel splitting is in its role in the successive
cancellation decoding procedure that is provided in
\cite{ArikanPolarCodes}. The error performance analysis of this
decoding procedure relies on the following two results:

\begin{theorem}[\cite{TelatarArikanPolarRate}] \label{theorem:polarizationRate}
{\em Let $p$ be a binary-input symmetric DMC with capacity $C(p)$,
and fix an arbitrary rate $R<C(p)$ and a positive constant $\beta <
\frac{1}{2}$. Then, there exists a sequence of information sets
$\mathcal{A}_n \subset [n]$, where $n=2^i$, $i\in\mathbb{N}$, such
that for large enough blocklengths $n$ the following properties are
satisfied:
\begin{enumerate}
\item Rate:
\begin{equation*}
|\mathcal{A}_n| \geq nR.
\end{equation*}
\item Performance: The Bhattacharyya parameters in
\eqref{equation:BhattacharyyaParameter} satisfy
\begin{equation*}
B(p_n^{(l)}) \leq 2^{-n^\beta}
\end{equation*}
for every $l \in \mathcal{A}_n$.
\end{enumerate}
}
\end{theorem}

\begin{proposition}[\cite{ArikanPolarCodes}] \label{propositon:errorBoundPolar}
{\em Assume that the vector $\mathbf{w} = (w_1,\dots,w_n) \in
\mathcal{X}^n$ is encoded via the considered recursive construction
in \eqref{equation:equivalentRecursiveConstruction}, and is
transmitted over a memoryless and symmetric DMC channel $p$ with a
binary-input alphabet $\mathcal{X}$ and an output alphabet
$\mathcal{Y}$. Define the event
\begin{equation} \label{equation:eventDefinitionErrorInL}
\mathcal{E}_l(p) \triangleq \left\{ p_n^{(l)}(\mathbf{y},
\mathbf{w}^{(l-1)}|w_l) \leq p_n^{(l)}(\mathbf{y},
\mathbf{w}^{(l-1)}|w_l+1)\right\}
\end{equation}
where $\mathbf{y}\in \mathcal{Y}^n$ is the received vector,
$\mathbf{w}^{(l-1)} = (w_1,\ldots,w_{l-1})$ is the vector comprises
of the first $l-1$ bits of $\mathbf{w}$, $p_n^{(l)}$ is the split
channel in \eqref{equation:standradPolarChannelSplitting} and the
addition is carried modulo 2. Then, the event $\mathcal{E}_l$ is
independent with the actual input vector $\mathbf{w}$ and
\begin{equation*}
\Pr\bigl(\mathcal{E}_l(p)\bigr) \leq B\bigl(p_n^{(l)}\bigr)
\end{equation*}
where $B(p_n^{(l)})$ is the Bhattacharyya parameter in
\eqref{equation:BhattacharyyaParameter}.}
\end{proposition}

\section{The Proposed Scheme: The Physically Degraded Case}
\label{section:TheProposedScheme}

\subsection{Polar Coding for Degraded Wire-Tap Channels}
\label{section:polarCoding4DegradedWireTapChannel}

\subsection*{Coset Block Codes}

A polar coding scheme is defined for the wire-tap channel. The
proposed scheme is defined using the notion of coset block codes,
based on the polar generator matrix $G_N$ introduced in
Section~\ref{section:polarCodes}. For a given block length $n=2^i$,
$i\in \mathbb{N}$, let $\mathcal{A}_n$ be an arbitrary subset of
$[n]$ of size $k$. In addition, let $\mathcal{N}_n$ be an additional
arbitrary subset of $\mathcal{A}_n^{\text{c}}$, of size $k^*$, and let $\mathbf{b}_n \in \mathcal{X}^{n-k
-k^*}$ be a length $n-k-k^*$ binary vector. Denote by
$\mathcal{B}_n$ the set of remaining indices in
$\mathcal{A}_n^{\text{c}}$, that is
\begin{equation} \label{equation:TheFixedBitsIndices}
\mathcal{B}_n \triangleq \mathcal{A}_n^{\text{c}} \setminus
\mathcal{N}_n.
\end{equation}
The sets $\mathcal{A}_n$, $\mathcal{B}_n$, and $\mathcal{N}_n$, the
polar generator matrix $G_n$ and the vector $\mathbf{b}_n$ are all
known to both the legitimate user and the eavesdropper.

Let $\mathbf{u} \in \mathcal{X}^k$ be a confidential information bit
vector that needs to be transmitted to the legitimate user. The
operation of the proposed secrecy scheme is described as follows:
\begin{enumerate}
\item A binary vector $\mathbf{b}_n^* \in
\mathcal{X}^{k^*}$ is chosen uniformly at random.
\item The coset block code $\mathcal{C}^*_n$ is chosen according to
\begin{equation} \label{equation:ChossingCosetBlockCode}
\mathcal{C}^*_n = \mathcal{C}\bigl(G_n(\mathcal{A}_n), \mathbf{b}_n
G_n(\mathcal{B}_n) + \mathbf{b}_n^*G_n(\mathcal{N}_n)\bigr).
\end{equation}
\item The information vector $\mathbf{u}$ is encoded into a codeword $\mathbf{x}$
using the coset block code $\mathcal{C}^*_n$. That is,
\begin{equation} \label{equation:scureCodeword}
\mathbf{x} = \mathbf{u}G_n(\mathcal{A}_n) + \mathbf{b}_n
G_n(\mathcal{B}_n) + \mathbf{b}_n^*G_n(\mathcal{N}_n)
\end{equation}
and it is transmitted over the wire-tap channel.
\end{enumerate}

If the complexity of constructing a random vector is considered as
$\mathcal{O}(1)$, then the encoding complexity of the proposed
scheme equals the encoding complexity of the single-user polar
encoding in~\cite{ArikanPolarCodes}, which is~$\mathcal{O}(n\log
n)$.

For given sets $\mathcal{A}_n$ and $\mathcal{N}_n$, and a vector
$\mathbf{b}_n$, the resulting coding scheme is denoted by
$\mathcal{C}_n(\mathcal{A}_n,\mathcal{N}_n,\mathbf{b}_n)$. Since
symmetric channels are considered, the performance of the provided
scheme is shown in the following to be independent with the actual
choice of $\mathbf{b}_n$. Consequently, the suggested coding scheme
is denoted by $\mathcal{C}_n(\mathcal{A}_n,\mathcal{N}_n)$.

\subsection*{Recursive Polar Construction}
\label{section:RecursivePolarConstruction} An equivalent recursive
construction of the proposed scheme is provided. Similarly to the
single-user construction
in~\eqref{equation:FirstStepRecursiveConstructionOfPolarCodes}, the
first step of the recursive construction is the composition of the
wiretap channel $P_2$ with an input alphabet from $\mathcal{X}^2$
and an output alphabet from $\mathcal{Y}^2 \times \mathcal{Z}^2$
\begin{equation}
\label{equation:FirstStepRecursiveConstructionOfSecurePolarCodes}
P_2(y_1,y_2,z_1,z_2|w_1,w_2) = P(y_1,z_1|w_1 + w_2) P(y_2,z_2|w_2)
\end{equation}
where $(y_1,y_2)\in\mathcal{Y}^2$, $(z_1,z_2)\in\mathcal{Z}^2$,
$(w_1,w_2)\in\mathcal{X}^2$, and the addition is carried modulo 2.

The continuation of the recursive construction follows in a similar
manner to the recursion in Section~\ref{section:polarCodes}; The
transition probability function $P_n$ for a channel with an input
alphabet $\mathcal{X}^n$ and an output alphabet $\mathcal{Y}^n
\times \mathcal{Z}^n$, is constructed using two independent copies
of a channel $P_{\frac{n}{2}}$ with an input alphabet
$\mathcal{X}^{\frac{n}{2}}$ and an output alphabet
$\mathcal{Y}^{\frac{n}{2}} \times \mathcal{Z}^{\frac{n}{2}}$. Note
that as in Section~\ref{section:polarCodes}, all block lengths ($n$)
are integral powers of 2. The first part of the recursive step
includes the evaluation of the vectors $\mathbf{s}, \mathbf{v} \in
\mathcal{X}^n$. This part is identical to the construction as
described in Section~\ref{section:polarCodes}
(steps~\ref{step:ConstructingTheSVectorInPolarizationCodes}
and~\ref{step:ConstructingTheVVectorInPolarizationCodes}). Finally,
the transition probability function $P_n(\mathbf{y}|\mathbf{x})$ is
given by
\begin{align}
P_n(\mathbf{y}, \mathbf{z}|\mathbf{x}) = & P_{\frac{n}{2}}
\left((y_1,y_2,\ldots,y_{\frac{n}{2}}),
(z_1,z_2,\ldots,z_{\frac{n}{2}})|(v_1,v_2,\ldots,v_{\frac{n}{2}})\right)
\nonumber\\
\label{equation:combinedWireTapChannel} & \cdot P_{\frac{n}{2}}
\left((y_{\frac{n}{2}+1},y_{\frac{n}{2}+2},\ldots,y_{n}),
(z_{\frac{n}{2}+1},z_{\frac{n}{2}+2},\ldots,z_{n})|(v_{\frac{n}{2}+1},v_{\frac{n}{2}+2},\ldots,v_{n})\right).
\end{align}
The channel $P_n$ in \eqref{equation:combinedWireTapChannel} is the
combined wire-tap channel.

As in the case of standard polar coding for the single-user model,
the recursive construction can be shown to be equivalent to a linear
encoding with the polar generator matrix $G_n$:

\begin{proposition} \label{proposition:RecursiveEquivalenceWireTapChannel}
{\em Let $P$ be a binary memoryless wire-tap channel with an input
alphabet $\mathcal{X}$ and output alphabets $\mathcal{Y}$ and
$\mathcal{Z}$, for the legitimate user and the eavesdropper,
respectively. In addition, let $P_n$ and $G_n$ be the combined
wire-tap channel in \eqref{equation:combinedWireTapChannel} and the
polar generator matrix in \eqref{equation:GRecursiveDefinition},
respectively. Then,
\begin{equation}
\label{equation:recrusiveEquivalence4ScurePolarCoding}
P_n(\mathbf{y},\mathbf{z}|\mathbf{w}) =
P(\mathbf{y},\mathbf{z}|\mathbf{w}G_n)
\end{equation}
for all $\mathbf{w} \in \mathcal{X}^n$,
$\mathbf{y}\in\mathcal{Y}^n$, and $\mathbf{z}\in\mathcal{Z}^n$.}
\end{proposition}

\begin{proof}
The proof of \eqref{equation:recrusiveEquivalence4ScurePolarCoding}
is identical to the proof of
\eqref{equation:equivalentRecursiveConstruction} in
\cite{ArikanPolarCodes}, where symbols from the output alphabet of
the single user channel are replaced with the corresponding pair of
symbols from the composite output alphabet (of the legitimate and
the eavesdropper channels).
\end{proof}

To obtain the equivalence of the recursive construction of the
combined channel $P_n$ in~\eqref{equation:combinedWireTapChannel}
with the encoding operation in~\eqref{equation:scureCodeword}, the
division of the components of $\mathbf{w}$ in
\eqref{equation:recrusiveEquivalence4ScurePolarCoding} for
information bits, random bits and predetermined and fixed bits, is
detailed. This division is defined by the sets $\mathcal{A}_n$ and
$\mathcal{N}_n$ as follows:
\begin{enumerate}
\item Over the indices specified by the index set $\mathcal{A}_n$,
the information bits $\mathbf{u}$ are placed.
\item The random bits
$\mathbf{b}_n^*$ are placed in the indices specified by
$\mathcal{N}_n$.
\item The predetermined and fixed bits in
$\mathbf{b}_n$ are left for the remaining indices specified by
$\mathcal{B}_n$.
\end{enumerate}
Plugging $\mathbf{u}$, $\mathbf{b}_n^*$, and $\mathbf{b}_n$ in
$\mathbf{w}G_n$, results in the coded message $\mathbf{x}$ in
\eqref{equation:scureCodeword}.

\subsection*{Channel Splitting and Degradation Properties}
The channel splitting operation in
\eqref{equation:standradPolarChannelSplitting} is repeated for the
case of wire-tap channels. This procedure can be carried in two
different but equivalent options:
\begin{enumerate}
\item First performing a channel splitting operation for the
wire-tap channel. This operation results in the split wire-tap
channels $\{P_n^{(l)}\}_{l=1}^n$ with a binary input alphabet
$\mathcal{X}$ and an output alphabet $\mathcal{Y}^n \times
\mathcal{Z}^n \times \mathcal{X}^{l-1}$:
\begin{equation} \label{equation:splittingOperation4WireTapChannel}
P_n^{(l)}(\mathbf{y}, \mathbf{z}, \mathbf{w}|w) \triangleq
\frac{1}{2^{n-1}} \sum_{\mathbf{c} \in \mathcal{X}^{n-l}}
P_n\bigl(\mathbf{y}, \mathbf{z} |(\mathbf{w},w,\mathbf{c})\bigr)
\end{equation}
where $\mathbf{y} \in \mathcal{Y}^n$, $\mathbf{z} \in
\mathcal{Z}^n$, $\mathbf{w} \in \mathcal{X}^{l-1}$, and $w \in
\mathcal{X}$. Next, deriving the marginal split channels
\begin{equation} \label{equation:splitLegitimateChannel}
G_n^{(l)}(\mathbf{y}, \mathbf{w}|w) \triangleq \sum_{\mathbf{z} \in
\mathcal{Z}^n} P_n^{(l)}(\mathbf{y}, \mathbf{z}, \mathbf{w}|w)
\end{equation}
and
\begin{equation} \label{equation:splitEavesdropperChannel}
Q_n^{(l)}(\mathbf{z}, \mathbf{w}|w) \triangleq \sum_{\mathbf{y} \in
\mathcal{Y}^n} P_n^{(l)}(\mathbf{y}, \mathbf{z}, \mathbf{w}|w)
\end{equation}
for the legitimate-user and eavesdropper, respectively, where
$\mathbf{y}$, $\mathbf{z}$, $\mathbf{w}$, and $w$ are as in
\eqref{equation:splittingOperation4WireTapChannel}.
\item First deriving the marginal combined channels:
\begin{equation} \label{equation:marginalLegitimateCombinedChannel}
G_n(\mathbf{y}|\mathbf{w}) \triangleq \sum_{\mathbf{z} \in
\mathcal{Z}^n} P_n(\mathbf{y},\mathbf{z}|\mathbf{w})
\end{equation}
and
\begin{equation} \label{equation:marginalEavesdropperCombinedChannel}
Q_n(\mathbf{z}|\mathbf{w}) \triangleq \sum_{\mathbf{y} \in
\mathcal{Y}^n} P_n(\mathbf{y},\mathbf{z}|\mathbf{w})
\end{equation}
for the legitimate user and eavesdropper, respectively, where
$\mathbf{y} \in \mathcal{Y}^n$, $\mathbf{z} \in \mathcal{Z}^n$, and
$\mathbf{w} \in \mathcal{X}^{n}$. Next, split the marginal combined
channels in \eqref{equation:marginalLegitimateCombinedChannel} and
\eqref{equation:marginalEavesdropperCombinedChannel} according to
\begin{equation} \label{equation:splitingLegitimateMarginal}
\frac{1}{2^{n-1}} \sum_{\mathbf{c} \in \mathcal{X}^{n-l}}
G_n\bigl(\mathbf{y} |(\mathbf{w},w,\mathbf{c})\bigr).
\end{equation}
and
\begin{equation} \label{equation:splitingEavesdropperMarginal}
\frac{1}{2^{n-1}} \sum_{\mathbf{c} \in \mathcal{X}^{n-l}}
Q_n\bigl(\mathbf{z} |(\mathbf{w},w,\mathbf{c})\bigr)
\end{equation}
where $\mathbf{y}$, $\mathbf{z}$, $\mathbf{w}$, and $w$ are as in
\eqref{equation:splittingOperation4WireTapChannel}.
\end{enumerate}
It is an immediate consequence of the equivalence properties
in~\eqref{equation:equivalentRecursiveConstruction}
and~\eqref{equation:recrusiveEquivalence4ScurePolarCoding}, that the
split channels in \eqref{equation:splitLegitimateChannel} and
\eqref{equation:splitEavesdropperChannel} equal to the channels in
\eqref{equation:splitingLegitimateMarginal} and
\eqref{equation:splitingEavesdropperMarginal}.

The following proposition considers physically degraded wire-tap
channels:
\begin{proposition}
\label{proposition:physicalDegradationOfSplitChannel} {\em Assume
that the wire-tap channel $P$ is physically degraded. Then, the
split channel $P_n^{(l)}(\mathbf{y}, \mathbf{z}, \mathbf{w}|x)$ in
\eqref{equation:splittingOperation4WireTapChannel} satisfies
\begin{equation} \label{equation:statementOfSplitDegradation}
P_n^{(l)}(\mathbf{y}, \mathbf{z}, \mathbf{w}|x) = G_n^{(l)}
(\mathbf{y}, \mathbf{w}|x) D(\mathbf{z}|\mathbf{y})
\end{equation}
where $G_n^{(l)}$ is the marginal split channel of the legitimate
user in \eqref{equation:splitLegitimateChannel}, $\mathbf{y} =
(y_1,\ldots,y_n) \in \mathcal{Y}^n$, $\mathbf{z} = (z_1,\ldots,z_n)
\in \mathcal{Z}^n$, $\mathbf{u} \in \mathcal{X}^{l-1}$, $x\in
\mathcal{X}$, $D(\mathbf{z}|\mathbf{y})$ is a memoryless transition
probability law:
\begin{equation*}
D(\mathbf{z}|\mathbf{y}) = \prod_{l=1}^n D(z_i|y_i)
\end{equation*}
and $D(z|y)$ is the conditional probability law of receiving a
symbol $z\in\mathcal{Z}$ at the eavesdropper, assuming that the
symbol $y\in \mathcal{Y}$ is received at the legitimate receiver.}
\end{proposition}

\begin{proof}
The recursion operation in
Proposition~\ref{proposition:splitRecursion} is valid for the
wire-tap channel. Specifically, for all $i>0$ and $1\leq l \leq 2^i$
it follows that
\begin{align}
\label{equation:oddTransformation4SplitWireTapChannel}
P_{2^{i+1}}^{(2l-1)}\bigl((\mathbf{y}^{(1)},\mathbf{y}^{(2)}),
(\mathbf{z}^{(1)},\mathbf{z}^{(2)}), \mathbf{w}|w_1\bigr) = &
\sum_{w \in \mathcal{X}} \frac{1}{2} P_{2^i}^{(l)}
\bigl(\mathbf{y}^{(1)},\mathbf{z}^{(1)}, g(\mathbf{w})|w_1 + w\bigr)
P_{2^i}^{(l)}\bigl(\mathbf{y}^{(2)},\mathbf{z}^{(2)},e(\mathbf{w})|w\bigr)\\
\label{equation:evenTransformation4SplitWireTapChannel}
P_{2^{i+1}}^{(2l)}\bigl((\mathbf{y}^{(1)},\mathbf{y}^{(2)}),(\mathbf{z}^{(1)},\mathbf{z}^{(2)}),(\mathbf{w},w_1)|w_2\bigr)
= & \frac{1}{2} P_{2^i}^{(l)}
\bigl(\mathbf{y}^{(1)},\mathbf{z}^{(1)},g(\mathbf{w})|w_1 +
w_2\bigr)
P_{2^i}^{(l)}\bigl(\mathbf{y}^{(2)},\mathbf{z}^{(2)},e(\mathbf{w})|w_2\bigr)
\end{align}
where $\mathbf{y}^{(1)}, \mathbf{y}^{(2)} \in \mathcal{Y}^{2^i}$,
$\mathbf{z}^{(1)}, \mathbf{z}^{(2)} \in \mathcal{Z}^{2^i}$,
$\mathbf{w} \in \mathcal{X}^{2l-2}$, $w_1, w_2 \in \mathcal{X}$, and
$g(\mathbf{w})$ and $e(\mathbf{w})$ are as defined
in~\eqref{equation:bitWizeOddEvenXorDefinition}
and~\eqref{equation:EvenVectorDefinition}, respectively. The proof
of the recursion property in
\eqref{equation:oddTransformation4SplitWireTapChannel} and
\eqref{equation:evenTransformation4SplitWireTapChannel} follows the
exact derivation as in~\cite{ArikanPolarCodes} (while replacing the
output alphabet of the single-user channel with the combined outputs
of the legitimate user and the eavesdropper).

From~\eqref{equation:marginalLegitimateCombinedChannel},
\eqref{equation:oddTransformation4SplitWireTapChannel},
and~\eqref{equation:evenTransformation4SplitWireTapChannel}, a
similar recursion follows for the marginal split channel
$G_n^{(l)}(\mathbf{y}, \mathbf{w}|x)$ of the legitimate user. To
this end, the recursion operations
in~\eqref{equation:oddTransformation4Splitchannel} and
\eqref{equation:evenTransformation4Splitchannel} are satisfied where
$p_{2^{i+1}}^{(2l-1)}$, $p_{2^i}^{(l)}$ and $p_{2^{i+1}}^{(2l)}$ are
replaced by $G_{2^{i+1}}^{(2l-1)}$, $G_{2^i}^{(l)}$ and
$G_{2^{i+1}}^{(2l)}$, respectively.

The proof of the degradation in
\eqref{equation:statementOfSplitDegradation} is accomplished by
induction. At the first step, from
\eqref{equation:oddTransformation4SplitWireTapChannel} and
\eqref{equation:evenTransformation4SplitWireTapChannel} it follows
that
\begin{align}
\label{equation:FirstOddSplitWireTapChannel}
P_{2}^{(1)}\bigl((y_1,y_2), (z_1,z_2) | w_1\bigr) = & \sum_{w \in
\mathcal{X}} \frac{1}{2} P
\bigl(y_1, z_1 | w_1 + w\bigr) P \bigl(y_2,z_2 | w\bigr)\\
\label{equation:FirstEvenSplitWireTapChannel}
P_{2}^{(2)}\bigl((y_1,y_2),(z_1,z_2),w_1|w_2\bigr) = & \frac{1}{2} P
\bigl(y_1,z_1|w_1 + w_2\bigr) P \bigl(y_2,z_2|w_2\bigr).
\end{align}
Then, plugging~\eqref{equation:degradationProperty}
in~\eqref{equation:FirstOddSplitWireTapChannel}
and~\eqref{equation:FirstEvenSplitWireTapChannel} concludes the
proof for the first step. Next, assume that the split channel
$P_{2^i}^{(l)}$ satisfies the degradation property
in~\eqref{equation:statementOfSplitDegradation}. That is, assume
that
\begin{equation} \label{equation:degradationAssumption4Induction}
P_{2^i}^{(l)}(\mathbf{y}, \mathbf{z}, \mathbf{w}'|w) = G_{2^i}^{(l)}
(\mathbf{y}, \mathbf{w}'|w) D(\mathbf{z}|\mathbf{y})
\end{equation}
for all $1\leq l \leq 2^i$, $\mathbf{y} \in \mathcal{Y}^{2^i}$,
$\mathbf{z} \in \mathcal{Z}^{2^i}$, $\mathbf{w}' \in
\mathcal{X}^{l-1}$, and $w\in\mathcal{X}$. Then,
from~\eqref{equation:oddTransformation4SplitWireTapChannel}
and~\eqref{equation:degradationAssumption4Induction} it follows that
\begin{align*}
P_{2^{i+1}}^{(2l-1)}\bigl((\mathbf{y}^{(1)},\mathbf{y}^{(2)}),
(\mathbf{z}^{(1)},\mathbf{z}^{(2)}), \mathbf{w}|w_1\bigr) = &
\sum_{w \in \mathcal{X}} \frac{1}{2} G_{2^i}^{(l)}
\bigl(\mathbf{y}^{(1)}, g(\mathbf{w})|w_1 + w\bigr)
D(\mathbf{z}^{(1)}|\mathbf{y}^{(1)})
\\
& \ \ \ \ \
G_{2^i}^{(l)}\bigl(\mathbf{y}^{(2)},e(\mathbf{w})|w\bigr)
D(\mathbf{z}^{(2)}|\mathbf{y}^{(2)})\\
= & G_{2^{i+1}}^{(2l-1)}\bigl((\mathbf{y}^{(1)},\mathbf{y}^{(2)}),
\mathbf{w}|w_1\bigr) \\
& \ \ \ \ \
D\bigl((\mathbf{z}^{(1)},\mathbf{z}^{(2)})|(\mathbf{y}^{(1)},\mathbf{y}^{(2)})\bigr)
\end{align*}
where the last step follows using the recursion properties of the
marginal split channel for the legitimate user. A similar argument
assures the degradation property for $P_{2^{i+1}}^{(2l)}$ which
concludes the proof of the proposition.
\end{proof}

\subsection*{Successive Cancellation Decoding}

The successive cancellation decoding procedure
in~\cite{ArikanPolarCodes} is applied for the legitimate user. The
difference from the standard single-user case is that for the
wire-tap channel model the legitimate user needs to decode both the
message $\mathbf{u} \in \mathcal{X}^k$ and the noisy vector
$\mathbf{b}_n^* \in \mathcal{X}^{k^*}$. In terms of information
sets, the legitimate receiver operates on the indices specified by
both $\mathcal{A}_n$ and $\mathcal{N}_n$. Denote by $\mathbf{w} =
(w_1,\ldots, w_n) \in \mathcal{X}^n$ the transmitted vector over the
combined channel $P_n$, then $\mathbf{w}$ is composed from the
information vector $\mathbf{u}$, the random vector $\mathbf{b}_n^*$,
and the predetermined fixed vector $\mathbf{b}_n$. It is important
not to confuse $\mathbf{w}$ with the actual codeword $\mathbf{x}$
in~\eqref{equation:scureCodeword}, which is transmitted over the
given wire-tap channel $P$. Both interpretations are equivalent as
the coset block code is equivalent to the recursive combining
construction. Nevertheless, the decoding rule (and its performance
analysis in the following) is characterized in terms of the vector
$\mathbf{w}$, transmitted over the combined wire-tap channel and
received over the marginal split channels for the legitimate user.

The decoding rule operates recursively to compute the length-$n$
decoded vector $\hat{\mathbf{w}} = (\hat{w}_1,\ldots, \hat{w}_n) \in
\mathcal{X}^n$. Let $1\leq l \leq n$, and assume that the first
$l-1$ components of $\hat{w}$, denoted by
$\hat{\mathbf{w}}^{(l-1)}$, are already evaluated. If $l \not \in
\bar{\mathcal{A}}_n$, where
\begin{equation*} \bar{\mathcal{A}}_n \triangleq \mathcal{A}_n
\cup \mathcal{N}_n.
\end{equation*}
then the current index $l$ is not in the information index set
$\mathcal{A}_n$ and not in the indices specified in $\mathcal{N}_n$
for the noisy vector. Consequently, $l \in \mathcal{B}_n$. Recall
that for the indices specified by $\mathcal{B}_n$, the predetermined
vector $\mathbf{b}_n$ is set. Since $\mathbf{b}_n$ is predetermined
and known (both to the legitimate user and the eavesdropper), $w_l$
is known at the receiver and therefore it is possible to set
\begin{equation*}
\hat{w}_l = w_l.
\end{equation*}
If $l \in \bar{\mathcal{A}}_n$, then the current index is identified
either as an information bit in $\mathbf{u}$ or as a noisy bit in
$\mathbf{b}^*_n$. For this case, the following decoding rule is
applied to the marginal split channel $G_n^{(l)}$ in
\eqref{equation:splitLegitimateChannel}:
\begin{equation} \label{equation:decodingRuleAtSplitChannel}
\hat{w}_l = \left\{
\begin{array}{cc} 0 & \text{if }
G_n^{(l)}(\mathbf{y}, \hat{\mathbf{w}}^{(l-1)}|0) \geq
G_n^{(l)}(\mathbf{y}, \hat{\mathbf{w}}^{(l-1)}|1) \\ 1 & \text{else}
\end{array}
\right..
\end{equation}

The successive cancellation decoding described in this section, is
by no mean optimal. This important observation is already noted for
the single-user case in~\cite{ArikanPolarCodes}. Nevertheless, for
an uncoded communication model with a communication channel whose
transition probability function is $G_n^{(l)}$, the detection rule
for the single bit $w_l$
in~\eqref{equation:decodingRuleAtSplitChannel} is optimal, if $w_l$
is an equiprobable bit.

\subsection{A Secrecy Achieving Property for Degraded Channels}
\label{section:SecrecyAchievingPropert}

\begin{theorem} \label{theorem:secrecyAchevingDegradedCase} {\em
Let $P$ be a binary-input, memoryless, degraded and symmetric
wire-tap channel with a secrecy capacity $C_{\text{s}}(P)$. Fix an
arbitrary positive $\beta < \frac{1}{2}$, and $R < C_{\text{s}}(P)$.
Then, there exist sequences of sets  $\mathcal{A}_n$ and
$\mathcal{N}_n$ such that the secrecy coding scheme
$\mathcal{C}_n(\mathcal{A}_n,\mathcal{N}_n)$ satisfies the following
properties:
\begin{enumerate}
\item Rate: For a sufficiently large block length $n$
\begin{equation} \label{eqaution:codeRatePropertyInTheorem}
R \leq \frac{1}{n}|\mathcal{A}_n|.
\end{equation}

\item Security: The equivocation rate
$R_{\text{e}}(\mathcal{C}_n(\mathcal{A}_n,\mathcal{N}_n)$ satisfies
\begin{equation} \label{equation:equivocationRatePropertyInTheorem}
\lim_{n\to\infty}
R_{\text{e}}\bigl(\mathcal{C}_n(\mathcal{A}_n,\mathcal{N}_n)\bigr)
\geq R.
\end{equation}
\item Reliability: The average block error probability under successive cancellation decoding
$P_{\text{e}}(\mathcal{C}_n(\mathcal{A},\mathcal{N}))$ satisfies
\begin{equation*}
P_{\text{e}}\bigl(\mathcal{C}_n(\mathcal{A}_n,\mathcal{N}_n)\bigr) =
o\left(2^{-n^\beta}\right).
\end{equation*}

\end{enumerate}}
\end{theorem}

\begin{proof}

The proof comprises of three parts: A code construction part where
the construction of the sets $\mathcal{A}_n$ and $\mathcal{N}_n$ is
described in details, along with the derivation of the coding rate
property in~\eqref{eqaution:codeRatePropertyInTheorem}. An analysis
of the equivocation rate is provided in the second part of the
proof. Finally, in the third part an upper bound on the block error
probability at the legitimate receiver is provided under successive
cancellation decoding.

\subsection*{Part I: The code construction}
Fix some $r^* = C(P_{Z|X}) - \epsilon$, and $r = C(P_{Y|X}) -
\epsilon$, where $C(P_{Y|X})$ and $C(P_{Z|X})$ are the channel
capacities of the marginal channels for the legitimate user and the
eavesdropper, and $\epsilon
> 0$ is determined later. According to Theorem~\ref{theorem:polarizationRate}, there exists a sequence of index
sets $\tilde{\mathcal{N}}_n \subset [n]$, satisfying:
\begin{enumerate}
\item The cardinality of the index set $\tilde{\mathcal{N}}_n$ satisfies
\begin{equation} \label{equation:CardinalityPropertyEavesdropper}
|\tilde{\mathcal{N}}_n| \geq \lfloor n r^* \rfloor.
\end{equation}
\item For all $l \in \tilde{\mathcal{N}}$, the Bhattacharyya parameter $B(Q_n^{(l)})$ of the split channel
$Q_n^{(l)}$ of the eavesdropper
in~\eqref{equation:splitEavesdropperChannel}, is upper bounded by
\begin{equation} \label{equation:BConstantPropertyEavesdropper}
B(Q_n^{(l)}) \leq 2^{-n^\beta}.
\end{equation}
\end{enumerate}
The index set $\mathcal{N}_n$ of size $\lfloor n r^* \rfloor$ is
chosen arbitrary from $\tilde{\mathcal{N}}_n$.

Next, Theorem~\ref{theorem:polarizationRate} is applied for the
marginal channel of the legitimate user. Accordingly, there exists a
sequence of index sets $\tilde{\mathcal{A}}_n \subset [n]$,
satisfying:
\begin{enumerate}
\item The cardinality of the index set $\tilde{\mathcal{A}}_n$ satisfies
\begin{equation} \label{equation:CardinalityPropertyLegitimate}
|\tilde{\mathcal{A}}_n| \geq \lfloor n r\rfloor.
\end{equation}
\item For all $l \in \tilde{\mathcal{A}}_n$, the Bhattacharyya parameter $B(G_n^{(l)})$ of the split channel
$G_n^{(l)}$ of the legitimate user
in~\eqref{equation:splitingLegitimateMarginal}, is upper bounded
according to
\begin{equation} \label{equation:BConstantPropertyLegitimate}
B(G_n^{(l)}) \leq 2^{-n^\beta}.
\end{equation}
\end{enumerate}
For each $n$, the information index set $\mathcal{A}_n$ of size
$\lfloor n r \rfloor - \lfloor n r^*\rfloor$ is chosen from
$\tilde{\mathcal{A}}_n \setminus \mathcal{N}_n$. As $|\mathcal{N}_n|
= \lfloor n r^*\rfloor$ and $|\tilde{\mathcal{A}}_n| \geq \lfloor n
r\rfloor$, the set $\tilde{\mathcal{A}}_n \setminus \mathcal{N}_n$
is of sufficient size. The specific choice of $\mathcal{A}_n$ may be
carried arbitrarily. Nevertheless, the best choice is to pick the
indices in $\tilde{\mathcal{A}}_n \setminus \mathcal{N}_n$ whose
corresponding marginal split-channels for the legitimate-user have
the lowest Bhattacharyya parameters.

The code rate of the resulting scheme satisfies
\begin{align}
\frac{1}{n}|\mathcal{A}_n| & \geq \frac{r-1}{n} -
\frac{r^*-1}{n} \nonumber\\
& = C(P_{Y|X}) - C(P_{Z|X}) - 2 \epsilon - \frac{2}{n} \nonumber\\
\label{equation:rateProofInequality} & = C_{\text{s}}(P) - 2\epsilon
-\frac{2}{n}
\end{align}
where the last equality follows from
Theorem~\ref{theorem:secrecyCapacity}. Consequently, for a large
enough block length and a properly chosen (small) $\epsilon$, the
code rate of the proposed scheme satisfies
\eqref{eqaution:codeRatePropertyInTheorem}.

The choice of the vector $\mathbf{b}_n \in \mathcal{X}^{n-k-k^*}$
may be carried arbitrarily.

\subsection*{Part II: The equivocation rate analysis}
The confidential message vector, the transmitted codeword, and the
received vector at the eavesdropper are denoted by the random
vectors $\mathbf{U}$, $\mathbf{X}$, and $\mathbf{Z}$, respectively.
The equivocation rate of the proposed scheme
$R_{\text{e}}\bigl(\mathcal{C}_n(\mathcal{A},\mathcal{N})\bigr)$ is
given by
\begin{align}
R_{\text{e}}\bigl(\mathcal{C}_n(\mathcal{A},\mathcal{N})\bigr) = &
\frac{1}{n} H(\mathbf{U}|\mathbf{Z}) \nonumber\\
& = \frac{1}{n}H(\mathbf{U}) - \frac{1}{n}I(\mathbf{U};\mathbf{Z}) \nonumber\\
\label{equation:eqovocationRateToBound} & = \frac{1}{n}
|\mathcal{A}_n|- \frac{1}{n}I(\mathbf{U};\mathbf{Z})
\end{align}
Where the last equality follows since the message bit vector is of
length $|\mathcal{A}_n|$ and equiprobable. Using the chain rule of
mutual information
\begin{align*}
I(\mathbf{U}, \mathbf{X}; \mathbf{Z}) = & I(\mathbf{U};\mathbf{Z}) +
I(\mathbf{X}; \mathbf{Z} | \mathbf{U}) \\
= & I(\mathbf{X};\mathbf{Z}) + I(\mathbf{U}; \mathbf{Z} |
\mathbf{X}).
\end{align*}
Consequently,
\begin{align}
I(\mathbf{U};\mathbf{Z}) = & I(\mathbf{X};\mathbf{Z}) +
I(\mathbf{U};
\mathbf{Z} | \mathbf{X}) - I(\mathbf{X}; \mathbf{Z} | \mathbf{U}) \nonumber\\
\stackrel{(a)} = & I(\mathbf{X};\mathbf{Z}) - I(\mathbf{X};
\mathbf{Z} | \mathbf{U}) \nonumber\\
\label{equation:mutualInformationInBound1} \leq & nC(P_{Z|X}) -
I(\mathbf{X}; \mathbf{Z} | \mathbf{U})
\end{align}
where (a) follows since $\mathbf{U} \rightarrow\mathbf{X}\rightarrow
\mathbf{Z}$ is a Markov chain which implies that $\mathbf{Z}$ and
$\mathbf{U}$ are statistically independent given $\mathbf{X}$, and
$C(P_{Z|X})$ is the channel capacity of the marginal channel to the
eavesdropper. The conditional mutual information $I(\mathbf{X};
\mathbf{Z} | \mathbf{U})$ is given by
\begin{align}
I(\mathbf{X}; \mathbf{Z} | \mathbf{U}) = & H(\mathbf{X}|\mathbf{U})
- H(\mathbf{X}|\mathbf{U}, \mathbf{Z}) \nonumber\\
\stackrel{(a)} = & |\mathcal{N}_n| - H(\mathbf{X}|\mathbf{U}, \mathbf{Z}) \nonumber\\
\label{equation:mutualInformationInBound2} \stackrel{(b)} \geq & n
\bigl(C(P_{Z|X})-\epsilon\bigr) - 1 - H(\mathbf{X}|\mathbf{U},
\mathbf{Z})
\end{align}
where (a) follows since the binary vector $\mathbf{b}^*$ is chosen
uniformly at random and it is independent with the confidential
message, and (b) follows since $|\mathcal{N}_n| = \lfloor n r^*
\rfloor$ and $r^* = C(P_{Z|X})-\epsilon$.

Let $P_{\text{e}|\mathbf{U}}$ denote the error probability of a
decoder that needs to decode $\mathbf{X}$ while having access to
both the eavesdropper observation vector $\mathbf{Z}$, the
confidential message vector $\mathbf{U}$, and the predetermined
vector $\mathbf{b}_n$ (which is fixed, predetermined, and known to
all the users in the model). Note that if both the confidential
message $\mathbf{U}$ and the predetermined vector $\mathbf{b}_n$ are
known at the receiver, then the remaining uncertainty in the
codeword $\mathbf{X}$ relates only to the random vector
$\mathbf{b}^*_n$ of size $\mathcal{N}_n$. Using Fano's inequality
(see, e.g., \cite{TomCovBook}), the conditional entropy
$H(\mathbf{X}|\mathbf{U}, \mathbf{Z})$ is bounded according to
\begin{align}
H(\mathbf{X}|\mathbf{U}, \mathbf{Z}) \leq & h_2
(P_{\text{e}|\mathbf{U}}) + P_{\text{e}|\mathbf{U}}
\log(2^{|\mathcal{N}_n|}-1) \nonumber\\
\label{equation:FanoAtEavesdropper} \leq & h_2
(P_{\text{e}|\mathbf{U}}) + n r^* P_{\text{e}|\mathbf{U}}
\end{align}
where $h_2(x) \triangleq -x\log x - (1-x) \log (1-x)$ is the binary
entropy function. from
\eqref{equation:eqovocationRateToBound}-\eqref{equation:FanoAtEavesdropper}
it follows that
\begin{align}
\label{equation:boundingTheEquivocationRate1}
R_{\text{e}}\bigl(\mathcal{C}_n(\mathcal{A},\mathcal{N})\bigr) \geq
& \frac{1}{n}|\mathcal{A}_n| - \epsilon - \frac{1}{n} - \frac{1}{n}
\left(h_2 \left( P_{\text{e}|\mathbf{U}} \right) + n r^*
P_{\text{e}|\mathbf{U}} \right) \\
\label{equation:boundingTheEquivocationRate} \geq & R - \frac{1}{n}
- \frac{1}{n} \left(h_2 \left( P_{\text{e}|\mathbf{U}} \right) + n
r^* P_{\text{e}|\mathbf{U}} \right)
\end{align}
where the last inequality follows
from~\eqref{equation:rateProofInequality} for a sufficiently small
$\epsilon$ and a sufficiently large $n$. The error probability
$P_{\text{e}|\mathbf{U}}$
in~\eqref{equation:boundingTheEquivocationRate} can be upper bounded
by the error probability under the suboptimal successive
cancellation decoder in~\cite{ArikanPolarCodes}, which is fully
informed with both the predetermined vector $\mathbf{b}_n$ and the
confidential message vector $\mathbf{U}$. It follows
from~\cite{TelatarArikanPolarRate} that
\begin{equation*}
P_{\text{e}|\mathbf{U}}  \leq o(2^{-n^\beta})
\end{equation*}
which concludes the proof of
\eqref{equation:equivocationRatePropertyInTheorem}.

\subsection*{Part III: The error performance at the legitimate decoder}

The successive cancellation decoding procedure at the legitimate
receiver is analyzed. First, fix a vector $\mathbf{w} =
(w_1,\ldots,w_n) \in \mathcal{X}^n$ comprises of the information
message $\mathbf{u} \in \mathcal{X}^k$, the randomly chosen vector
$\mathbf{b}^* \in \mathcal{X}^{k^*}$, and the predetermined vector
$\mathbf{b} \in \mathcal{X}^{n-k-k^*}$. The conditional block error
probability is denoted by $P_{\text{e}|\mathbf{w}}$. That is,
$P_{\text{e}|\mathbf{w}}$ is the probability of a block error event
given that the input vector is $\mathbf{w}$. Denote by
$\mathbf{w}^{(l)} = (w_1,\ldots,w_{l})$ the first $l$ bits of
$\mathbf{w}$, and by
$\hat{\mathbf{w}}^{(l)}=(\hat{w}_1,\ldots,\hat{w}_l)$ the first $l$
decoded bits. The event
\begin{equation*}
\mathcal{F}_l \triangleq \left\{\mathbf{w}^{(l-1)} =
\hat{\mathbf{w}}^{(l-1)},\ \ \ w_l \neq \hat{w}_l\right\}
\end{equation*}
corresponds to the case where the first $l-1$ bits of $\mathbf{w}$
are decoded correctly and the first decoding error is in the $l$-th
bit. Notice that
\begin{equation*}
\mathcal{F}_l \subset \mathcal{E}_l(G_n^{l})
\end{equation*}
where $\mathcal{E}_l$ is the event defined
in~\eqref{equation:eventDefinitionErrorInL}, and $G_n^{l}$ is the
marginal split channel in~\eqref{equation:splitLegitimateChannel}.
Consequently, it follows using the union bound that
\begin{align}
P_{\text{e}|\mathbf{w}} = & \Pr \bigl(\cup_{l=1}^n
\mathcal{F}_l|\ \mathbf{w}\bigr) \nonumber\\
\label{equation:unionBoundOnConditionalErrorEvent} & \leq \sum_{l
\in \bar{\mathcal{A}}_n} \Pr\bigl(\mathcal{E}_l(G_n^{(l)})|\
\mathbf{w} \bigr).
\end{align}

Next, the summation
in~\eqref{equation:unionBoundOnConditionalErrorEvent} is split to
two summations: a summation over the indices in $\mathcal{A}_n$ and
a summation over the indices in $\mathcal{N}_n$. For an index $l \in
\mathcal{A}_n$, it follows from
Proposition~\ref{propositon:errorBoundPolar} that for all
$\mathbf{w} \in \mathcal{X}^n$
\begin{equation} \label{equation:bound4InformationBitIndices}
\Pr\bigl(\mathcal{E}_l(G_n^{(l)})|\ \mathbf{w} \bigr) \leq
B(G_n^{(l)})
\end{equation}
where $B(G_n^{(l)})$ is the Bhattacharyya parameter in
\eqref{equation:BhattacharyyaParameter}. To address the probability
of the event $\mathcal{E}_l(G_n^{(l)})$ where $l \in \mathcal{N}_n$,
notice that at the output of the marginal split channel, the
decoding rule for $w_l$
in~\eqref{equation:decodingRuleAtSplitChannel} is
optimal\footnote{As stated, this optimality is only under the
setting of the split channel, and by no means implies optimality of
the complete procedure (which is clearly suboptimal).}. Recall the
degradation property in
Proposition~\ref{proposition:physicalDegradationOfSplitChannel}.
According to
Proposition~\ref{proposition:physicalDegradationOfSplitChannel} the
marginal split channel of the eavesdropper is physically degraded
with respect to the marginal split channel of the legitimate user.
Consequently, it is clearly suboptimal to first degrade the
observations at the split channel of the legitimate user, and only
then to detect the bit $w_l$ over the corresponding marginal split
channel of the eavesdropper. Specifically, $w_l$ is detected
according to
\begin{equation*}
\hat{w}_l = \left\{
\begin{array}{cc} 0 & \text{if }
Q_n^{(l)}(\mathbf{z}, \hat{w}^{(l-1)}|0) \geq Q_n^{(l)}(\mathbf{z},
\hat{w}^{(l-1)}|1) \\ 1 & \text{else}
\end{array}
\right.
\end{equation*}
where $\mathbf{z} \in \mathcal{Z}^n$ is a degraded version of
$\mathbf{y} \in \mathcal{Y}^n$, randomly picked according to the
probability law $D(\mathbf{z}|\mathbf{y})$ in
\eqref{equation:statementOfSplitDegradation}. This detection rule is
inferior with respect
to~\eqref{equation:decodingRuleAtSplitChannel}. Hence, based on
Proposition~\ref{propositon:errorBoundPolar}, the upper bound
\begin{equation} \label{equation:boundForRandomBitIndices}
\Pr\bigl(\mathcal{E}_l(G_n^{(l)})|\ \mathbf{w} \bigr) \leq
B(Q_n^{(l)})
\end{equation}
holds for all $l \in \mathcal{N}_n$. From
\eqref{equation:unionBoundOnConditionalErrorEvent},
\eqref{equation:bound4InformationBitIndices},
and~\eqref{equation:boundForRandomBitIndices}, it follows that the
average block error probability is upper bounded by
\begin{equation*}
P_{\text{e}}(\mathcal{C}_n(\mathcal{A},\mathcal{N})) \leq
\sum_{l\in\mathcal{A}_n} B(G_n^{(l)}) + \sum_{l\in\mathcal{N}_n}
B(Q_n^{(l)}).
\end{equation*}
The proof concludes using the bound on the polarization rate of the
Bhattacharyya parameter in Theorem~\ref{theorem:polarizationRate}
and the specific choice of the sets $\mathcal{A}_n$ and
$\mathcal{N}_n$.
\end{proof}

\begin{remark}[\textbf{On communicating with full capacity}] \label{remark:communicatingAtFullRate}
The noisy bits $\mathbf{b}^*_n$, defining the coset block code
$\mathcal{C}^*_n$ based on the noisy index set $\mathcal{N}_n$ (see
eq.~\eqref{equation:ChossingCosetBlockCode}), are reliably detected
by the legitimate user. It is therefore suggested to utilize these
bits in order to communicate with the legitimate user. That is,
instead of setting the bits in $\mathbf{b}^*_n$ to noisy random
bits, non-secret information bits are suggested to be set on
$\mathbf{b}^*_n$. The non-secret information bits must be
statistically independent and equiprobable. In addition, the
non-secret information must be statistically independent with the
secret-information. These statistical properties allows the
non-secret information bits to act as if they are noisy bits (where
the eavesdropper is concerned). As a result of the cardinality of
the index set
$\mathcal{A}_n$~\eqref{equation:CardinalityPropertyLegitimate}, the
overall rate, including secret and non-secret information, is
arbitrarily close the full (marginal) channel capacity of the
legitimate user $C(P_{Y|X})$.
\end{remark}

\begin{remark}[\textbf{The noisy bits must not be fixed}]
\label{remark:NosiyBitsCantBeFixed} It is important to note that the
bits in $\mathbf{b}^*_n$ must be chosen at random for each block
transmission. To see this, first note (based on the data processing
inequality) that
\begin{equation}
\label{equation:Inequality4MutualInformationOfNoisyBits} \frac{1}{n}
I(\mathbf{b}^*_n; \mathbf{Z}) \leq \frac{1}{n}
I(\textbf{X};\textbf{Z})
\end{equation}
for all $n>0$. Assuming
that~\eqref{equation:Inequality4MutualInformationOfNoisyBits} is
satisfied with equality. It follows that both the legitimate user
and the eavesdropper can reliably decoded the vector
$\mathbf{b}^*_n$. Considering the current setting as if it is a
broadcast communication problem over the given channel, a broadcast
scheme is therefore provided where we can reliably communicated with
the legitimate user at a rate arbitrarily close to its marginal
capacity $C(P_{Y|X})$ and at the same time with the eavesdropper at
a (common) rate which is arbitrarily close to
$\frac{1}{n}I(\mathbf{X};\mathbf{Z})$. This violates the fundamental
limit imposed by the capacity region of the degraded broadcast
channel (see, e.g.,~\cite{TomCovBook}). Consequently, it follows
that
\begin{equation}
\label{equation:StrictInequalityIfMutualIfnroamtions} \frac{1}{n}
I(\mathbf{b}^*_n; \mathbf{Z}) < \frac{1}{n} I(\textbf{X};\textbf{Z})
\end{equation}
for all $n>0$. Next, since there is a one-to-one correspondence
between the transmitted codeword $\mathbf{X}$ and the vector pair
which is comprised of the random bits $\mathbf{b}^*$ and the
confidential message $\mathbf{U}$ (the vector $\mathbf{b}$ is
predetermined and fixed), it follows that
\begin{align}
\nonumber
\frac{1}{n} I(\mathbf{X}; \mathbf{Z}) & = \frac{1}{n} I(\mathbf{U},\mathbf{b}^*;\mathbf{Z}) \\
\label{equation:mutualInformationEquality} & \stackrel{(\text{a})} =
\frac{1}{n} I(\mathbf{b}^*; \mathbf{Z}) + \frac{1}{n} I(\mathbf{U};
\mathbf{Z} | \mathbf{b}^*)
\end{align}
for all $n>0$, where (a) follows by the chain rule of mutual
information. Hence it is observed
from~\eqref{equation:StrictInequalityIfMutualIfnroamtions}
and~\eqref{equation:mutualInformationEquality} that
\begin{equation*}
\frac{1}{n} I(\mathbf{U}; \mathbf{Z} | \mathbf{b}^*) > 0
\end{equation*}
for all $n$. This assures that if the vector $\mathbf{b}^*$ is known
to the eavesdropper, for example by choosing a fixed $\mathbf{b}^*$,
perfect secrecy can not be established, not even in the weak sense.
\end{remark}

%
%IX;Y) = Rc + Rs
%
%Rs = I(X;Y) - I(X;Z)
%
%Rc = I(X;Z).
%
%Now as said:
%
%I(S,C;Z) = I(X;Z) = I(S;Z) + I(C;Z|S).
%
%Since I(S;Z) = 0, then I(C;Z|S) = I(X;Z).
%
%Now the other decomposition:
%
%IX;Z) = I(S,C;Z) = I(C;Z) + I(S;Z|C) = I(C;Z) + I(S;Z,C)
%
%(as S and C are independent).
%
%Now suppose:
%
%I(C;Z) =  Rc = I(X;Z).
%
%That means that the degraded channel was able to decode at rate Rc,
%and the better channel was able to decode at rate Rs+Rc (forget even
%secrecy). This is in violation to the capacity region of the
%degraded broadcast channel!. The capacity region of the degraded
%broadcast channel dictates then, if:
%
%Rs+Rc = I(X;Y),
%
%then no rate can be decoded at the degraded channel, irrelevant of
%secrecy. Thus:
%
%I(C;Z) < I(X;Z),
%
%and hence:
%
%I(S;Z,C) > 0 !

It is observed in~\cite{CKBook}, that if $(R_1,R_1)$ is an achievable rate-equivocation pair and in addition, an additional information rate $R_2$ is achievable without secrecy (that is, in the ordinary notion of reliable communication), then the  $(R_1+R_2, R_1)$ is also an achieved rate-equivocation pair. The other direction is also provided in~\cite[p. 411]{CKBook}. Following Remark~\ref{remark:communicatingAtFullRate} which suggests the option of communicating in full rate, and the observations in~\cite{CKBook}, it is expected that the entire rate-equivocation region is obtained with polar coding. This result is provided in the following corollary:

\begin{corollary}[\textbf{The entire rate-equivocation region is achievable with polar codes}] \label{corollary:EntireRegion}
{\em Under the assumptions and notation in Theorem~\ref{theorem:secrecyAchevingDegradedCase}, the entire rate-equivocation region is achievable with polar coding.}
\end{corollary}
\begin{proof} Take a rate-equivocation pair $(R,R_{\text{e}})$ in the rate-equivocation region defined in~\eqref{equation:rateEquivocationRegion}. Define $R_1 = R_{\text{e}}$, and $R_2 = R-R_1$. Note that $R_2 \geq 0$ as $R_{\text{e}} \leq R$. Consider the coset block code in~\eqref{equation:ChossingCosetBlockCode}. Since $R_{\text{e}} \leq C_s(P)$, the rate $R_1$ is achievable via the index set $\mathcal{A}_n$. It is further detailed in the proof of Theorem~\ref{theorem:secrecyAchevingDegradedCase}, that the information transmitted via the indices in $\mathcal{A}_n$ is secure. Specifically, it follows from~\eqref{equation:boundingTheEquivocationRate1} that the equivocation rate is arbitrarily close to $\frac{1}{n}|\mathcal{A}_n|$. As explained in Remark~\ref{remark:communicatingAtFullRate}, reliable communication (not necessarily secure) of an additional rate of up to the capacity $C(P_{Y|X})$ of the marginal channel to the legitimate user, is achievable. Therefore, the additional rate $R_2$, is achievable either via the remaining indices in $\mathcal{A}_n$ and the vector $\mathbf{b}^*_n$ corresponding to the indices in $\mathcal{N}_n$.
\end{proof}

\subsection{Secrecy Achieving Properties for Erasure Wiretap Channels}
\label{section:SecrecyAchievingPropertBEWC}

In this section, a particular case of binary erasure wiretap channel is considered. Specifically, it is assumed that the channel to the legitimate user is noiseless, and the channel to the eavesdropper is a binary erasure channel (BEC) with an erasure probability $\delta$, is considered.
%It is shown that for this particular case the proposed scheme achieves the secrecy capacity with the strong notion of secrecy.
Recall that the set sequence $\mathcal{N}_n$ of the indices that correspond to ``good'' split channel to the eavesdropper, is chosen as to achieve the capacity to the eavesdropper. As the channel to the legitimate user is noiseless, that is $\mathbf{y} = \mathbf{x}$, the set sequence $\mathcal{A}_n$ and is set according to
\begin{equation} \label{equation:informationSetBEC}
\mathcal{A}_n \triangleq [n] \setminus \mathcal{N}_n.
\end{equation}
Note that for this particular case $\mathcal{B}_n = \emptyset$. The resulting coding scheme is then a particular case of the coset coding scheme in~\cite{OzWyner} where the base code is determine by the generator matrix $G_n(\mathcal{N}_n)$ and the actual coset is determined by $\mathbf{u}G_n(\mathcal{A}_n)$ where $\mathbf{u}$ is the transmitted information bits (the secret message) and $G_n$ is the polar generator matrix for a block length $n$. Specifically, the codeword $\mathbf{x}$ is given, based on\eqref{equation:scureCodeword}, by
\begin{equation} \label{scureCodewordErasureChannel}
\mathbf{x} = \mathbf{u}G_n(\mathcal{A}_n) + \mathbf{b}_n^*G_n(\mathcal{N}_n).
\end{equation}

The rate and reliability properties in this particular case follows immediately as a result of Theorem~\ref{theorem:secrecyAchevingDegradedCase}. That is, the rate approaches the secrecy capacity, which in this case equals $\delta$, and the legitimate user obviously can decode the transmitted message. %It remains the show that a strong notion of secrecy is guaranteed.
As in the second part of the proof of Theorem~\ref{theorem:secrecyAchevingDegradedCase}, the confidential message vector, the transmitted codeword, and the
received vector at the eavesdropper are denoted by the random
vectors $\mathbf{U}$, $\mathbf{X}$, and $\mathbf{Z}$, respectively. The following lemma address the entropy measure $H(\mathbf{U}|\mathbf{Z})$.

\begin{lemma} \label{lemma:Strong} {\em Under the assumption and notation for the consider binary erasure wiretap channel, the entropy  $H(\mathbf{U}|\mathbf{Z})$ satisfies
\begin{equation*}
H(\mathbf{U}|\mathbf{Z}) \geq n \delta (1-c2^{-n^{\beta}})
\end{equation*}
where $\delta$ is the erasure probability of the wiretap channel, and $c>0$.}
\end{lemma}

\begin{proof} Let us fix a particular realization of the channel erasure sequence~\footnote{This case is studied in~\cite{OzWyner}, and some parts of the provided proof are based on proper presentation of the techniques developed in~\cite{OzWyner} for the case at hand.}. Denote by $\mathcal{D}$ the set of $\mu$ indices which are not erased. That is, the eavesdropper received the bits $X_i$ for every $i\in\mathcal{D}$, and erasure symbols for every index in $\mathcal{D}^{\text{c}} \triangleq [n]\setminus\mathcal{D}$. Consider the $|\mathcal{N}_n| \times n$ matrix $\{G_n(\mathcal{N}_n)\}$. As the generator matrix $G_n$ for the polar construction has a full rank (for every $n$ in the construction), the matrix $G_n(\mathcal{N}_n)$ has a rank $\mathcal{N}_n$. Therefore, it is a generator matrix for a binary linear block code of dimension $\mathcal{N}_n$. This code has a parity check matrix of size $|\mathcal{A}_n| \times n$, denoted by $H_n$ (recall that $\mathcal{A}$ is given by \eqref{equation:informationSetBEC}). Since all the information bits are equiprobable, and all the noisy bits are also equiprobable, the codeword $\mathbf{X}$, given by \eqref{equation:scureCodeword}, id uniformly distributed over all possible binary vectors in $\{0,1\}^n$. Consequently, all the bits in $\mathbf{X}$ are independent and identically distributed uniform binary random variables. Hence, $H(\mathbf{X}|\mathbf{Z}) = n-\mu$. In addition, note that if the codeword $\mathbf{X}$ is known, then information bits $\mathbf{U}$ are fully determined for the considered polar coding scheme. It follows that
\begin{align}
\label{equation:step5InWyner}
H(\mathbf{U}|\mathbf{Z}) & = H(\mathbf{U}|\mathbf{X},\mathbf{Z}) + H(\mathbf{X}|\mathbf{Z}) - H(\mathbf{X}|\mathbf{U},\mathbf{Z})\\
\label{equation:step6InWyner}
& = m-\mu-H(\mathbf{X}|\mathbf{U},\mathbf{Z}).
\end{align}
Note that \eqref{equation:step5InWyner} is a restatement of \cite[Eq.~(5)]{OzWyner}, and \eqref{equation:step6InWyner} is a restatement of \cite[Eq.~(6)]{OzWyner}.

Next, fix a realization $\mathbf{Z} = \mathbf{z} \in\{0,1\}^n$ and $\mathbf{U}=\mathbf{u} \in \{0,1\}^{|\mathcal{A}_n|}$. From~\eqref{scureCodewordErasureChannel}, it follows that the erased bits $\{X_i\}_{i\in\mathcal{D}^{\text{c}}}$ satisfies the linear equations
\begin{equation} \label{equation:erasureEquation}
\sum_{i\in\mathcal{D}} X_i \left(H_n\right)_i = H_n \mathbf{u} G_n\left(\mathcal{A}_n\right) + \sum_{i \in \mathcal{D}^{\text{c}}} X_i \left(H_n\right)_i
\end{equation}
where $\left(H_n\right)_i$ is the $i$-th column of the parity check matrix $H_n$. The number of solutions to~\eqref{equation:erasureEquation} is given by
\begin{equation*}
2^{n-\mu-d\left(\left\{\left(H_n\right)_i\right\}_{i\in\mathcal{D}}\right)}
\end{equation*}
where $d\left(\left\{\left(H_n\right)_i\right\}_{i\in\mathcal{D}}\right)$ is the dimension of the linear space spanned by the the column vectors in $\left\{\left(H_n\right)_i\right\}_{i\in\mathcal{D}}$. Since all the solutions for the erasures $X_i$, $i\in\mathcal{D}$, are equally likely, it follows that
\begin{equation} \label{equation:strongProofStep1}
H(\mathbf{X}|\mathbf{U}=\mathbf{u},\mathbf{Z}=\mathbf{z}) = n-\mu-d\left(\left\{\left(H_n\right)_i\right\}_{i\in\mathcal{D}}\right).
\end{equation}
From~\eqref{equation:step6InWyner} and~\eqref{equation:strongProofStep1}, it follows that
\begin{equation}
H(\mathbf{U}|\mathbf{Z}) = {\sf E} d\left(\left\{\left(H_n\right)_i\right\}_{i\in\mathcal{D}}\right).
\end{equation}
As the information indices $\mathcal{N}_n$ for the eavesdropper are chosen such that it can decode the noisy bits $\mathbf{b}^*$ with an error probability of $\mathcal{O}(2^{-n^\beta})$, it follows that
\begin{align}
H(\mathbf{U}|\mathbf{Z}) & \geq {\sf E}\left(d\left(\left\{\left(H_n\right)_i\right\}_{i\in\mathcal{D}}\right)|, \text{correct decoding}\right) (1-c2^{-n^{\beta}}) \\
& = n \delta (1-c2^{-n^{\beta}})
\end{align}
where $c>0$ and $\delta$ is the erasure probability of the eavesdropper channel.
\end{proof}

\begin{remark}[\textbf{All coset must be equally likely}]
 In the current discussion, the secrecy polar coding scheme is applied with $\mathcal{B}_n = \emptyset$. This fact is crucial for the proof of Lemma~\ref{lemma:Strong}. It is conjectured that this choice may be crucial to achieve the entire secrecy capacity under the strong secrecy condition.
\end{remark}

\begin{remark}[\textbf{On possible stronger notion of secrecy}]
Consider the conditions in Theorem~\ref{theorem:polarizationRate}. In particular, not that the rate $R<C(p)$ is kept fixed for the polarization structure of the code. If, it be possible to construct the sequence of polar codes, with a sequence of blocklength dependent rates $R_n$ having the property that
\begin{equation} \label{equation:rateSeq}
R_n \geq C(p)-\frac{\alpha}{n^\gamma}
\end{equation}
where $\alpha>0$ and $\gamma>1$ are arbitrarily fixed parameters. Then, it will follow as a corollary of Lemma~\ref{lemma:Strong} that a strong notion of secrecy is guaranteed. That is, the entropy $H(\mathbf{U}|\mathbf{Z})$ is arbitrarily close to $H(\mathbf{U})$. To see this, note that if polarization is possible while satisfying \eqref{equation:rateSeq}, it follows that
\begin{equation*}
|\mathcal{N}_n| \geq  n\left(1-\delta-\frac{\alpha}{n^\gamma}\right).
\end{equation*}
Consequently,
\begin{equation*}
H(\mathbf{U})=|\mathcal{A}_n|= n-|\mathcal{N}_n|=n \delta + \frac{\alpha}{n^{1-\gamma}}.
\end{equation*}
Hence $H(\mathbf{U}|\mathbf{Z})$ is lower bounded by a quantity which is arbitrarily close $H(\mathbf{U})$ as the blocklength increases. For the particular case of the BEC, it follows from~\cite[Eq. (34)-(35)]{ArikanPolarCodes}, that the considered question requires the analysis of the following sequence
\begin{equation*}
|\{i\in[n]:\ Z_n^i\leq C e^{\-n^{\beta}}\}|
\end{equation*}
where $\{Z_n^{i}\}_{i\in[n]}$ is a sequence, generated recursively according to
\begin{align*}
&Z_{2k}^{(2i-1)} = 2 Z_{k}^{(i)} - \left(Z_{k}^{(i)}\right)^2\\
&Z_{2k}^{(2i)} = \left(Z_{k}^{(i)}\right)^2.
\end{align*}
where $i\in[k]$ and $Z_1^{(1)} = \delta$.
\end{remark}

\section{An open polarization problem and the general wiretap channel}
\label{section:proposedScheme4GeneralWireTapeCh}

An open polarization problem is presented in addition to a
conjecture which suggests a possible solution. A polar secrecy
scheme for non-degraded wiretap channels is provided based on
suggested conjecture.

\subsection{On the polarization of the `bad' indices}
Let $\mathbf{W} = (W_1,\ldots,W_n)$ be a random vector, where
$\{W_i\}_{i=1}^n$ are statistically independent and equiprobable
$\Pr(W_i = 0) = \Pr(W_i = 1) = \frac{1}{2}$ for all $i\in[n]$. The
random vector $\mathbf{W}$ is polar encoded to a codeword
$\mathbf{X} = G_n \mathbf{W}$, where $G_n$ is the polar generator
matrix of size $n$. The codeword $\mathbf{X}$ is transmitted over a
binary input DMC $p$, whose output alphabet is $\mathcal{Y}$. The
received vector is denoted by $\mathbf{Y} = (Y_1,\ldots, Y_n)$. For
a given vector $\mathbf{W}$ and a set $\mathcal{A} \subseteq [n]$,
the following notation is used
\begin{equation*}
\mathbf{W}_{\mathcal{A}} \triangleq (W_{i_1},\ldots
W_{i_{|\mathcal{A}|}})
\end{equation*}
where $i_1 < i_2 < \ldots < i_{|\mathcal{A}|}$ and $i_{k} \in
\mathcal{A}$ for all $k \in [|\mathcal{A}|]$. Define the following
quantities of mutual information
\begin{equation} \label{equation:mutualInformationQuantities}
I_i \triangleq I(W_i; \mathbf{W}_{[i-1]}, \mathbf{Y}),\ \ i \in [n].
\end{equation}

The following polarization of mutual information is the key result
in~\cite{ArikanPolarCodes},~\cite{TelatarArikanPolarRate}:

\begin{theorem}[\textbf{On the polarization of mutual information \cite{ArikanPolarCodes}}] \label{theorem:PolarizationOfMutualInformation}
{\em Assume that $p$ is a binary-input output-symmetric DMC whose
capacity is $C(p)$, and fix $0< \delta < 1$. Then,
\begin{align*}
& \lim_{n\to\infty} \left(\frac{1}{n} \Bigl|\bigl\{i \in [n]:\ I_i
\in
(1-\delta,1]\bigr\}\Bigr|\right) = C(p) \\
& \lim_{n\to\infty} \left(\frac{1}{n} \Bigl|\bigl\{i \in [n]:\ I_i
\in [0, \delta) \bigr\}\Bigr|\right) = 1-C(p).
\end{align*}}
\end{theorem}

\vspace*{1cm}

Denote by $\mathcal{A}_n$ the set of indices for which the
corresponding mutual information quantities $I_i$,
$i\in\mathcal{A}_n$, are arbitrarily close to 1 bit (for a
sufficiently large $n$). The set $\mathcal{A}_n$ is called the
information index set. This is the very same index set in
Theorem~\ref{theorem:polarizationRate}, of `good' split channels
whose corresponding Bhattacharyya constants approach~0. Let
$\mathcal{A}'_n \subset \mathcal{A}_n$ and let $\mathcal{S}_n
\subseteq \mathcal{A}_n^{\text{c}}$. We define the index sets
\begin{equation*}
\mathcal{D}_n \triangleq \mathcal{A}'_n \cup \mathcal{S}_n
\end{equation*}
and
\begin{equation*}
\mathcal{D}^{(i)}_n \triangleq \bigl\{ j\in D_n:\ j < i\bigr\},\ \ i
\in [n].
\end{equation*}
A problem of interest lies in the $|\mathcal{D}_n|$ quantities of
mutual information:
\begin{equation} \label{equation:newMutualInformationQuantities}
J_i \triangleq I(W_i; \mathbf{W}_{\mathcal{D}^{(i)}_n},
\mathbf{W}_{\mathcal{D}^{\text{c}}_n}, \mathbf{Y}),\ \ \ i \in
\mathcal{D}_n.
\end{equation}
For the indices in $\mathcal{A}'_n$ a straight froward answer is
provided:

\begin{lemma}[\textbf{on the indices of `good' split channels}]
\label{lemma:OnModifiedPolarizationOfGoodBits} {\em Fix a
$0<\delta<1$ and an index $i\in\mathcal{A}'_n$. For sufficiently
large $n$
\begin{equation*}
J_i \geq 1-\delta.
\end{equation*}}
\end{lemma}
\begin{proof}
As the mutual information $I_i$
in~\eqref{equation:mutualInformationQuantities} includes a subset of
the random variables in $J_i$
in~\eqref{equation:newMutualInformationQuantities}, it follows that
\begin{equation*}
J_i \geq I_i.
\end{equation*}
The proof concludes using
Theorem~\ref{theorem:PolarizationOfMutualInformation} as
$\mathcal{A}'_n \subset \mathcal{A}_n$.
\end{proof}

According to Lemma~\ref{lemma:OnModifiedPolarizationOfGoodBits}
`good' indices for which the mutual information quantities $I_i$
approach 1 bit, remain `good'  with respect to the mutual
information $J_i$. The characterization of the 'bad' indices seems
at this point to be a greater challenge. A conjecture for possible
polarization properties of the mutual information quantities $J_i$
in~\eqref{equation:newMutualInformationQuantities} is provided for
the (`bad') indices in $\mathcal{S}_n$. Two possible polarization
properties are considered:

\begin{conjecture}[\textbf{On possible polarization dichotomy}] \label{conjecture:OnTheBadChannels}
{\em Fix a $0<\delta<1$. There exists a partition of $\mathcal{S}_n$
to two sets $\mathcal{S}'_n$ and $\mathcal{S}''_n = \mathcal{S}_n
\setminus \mathcal{S}'_n$, such that for a sufficiently large $n$
\begin{align}
\label{equation:conditionOnSTag}
& J_i < \delta,\ \ \text{for all } i \in S'_n \\
\label{equation:conditionOnSTagim} & J_i > 1-\delta,\ \ \text{for
all } i \in S''_n.
\end{align}
}
\end{conjecture}

%one of the following two options follows:
%\begin{enumerate}
%\item The same polarization of the mutual information quantities $I_i$ applies for the mutual information
%quantities $J_i$ for all $i\in \mathcal{S}_n$:
%\begin{equation*}
%J_i < \delta, \ \ \text{for all } i\in \mathcal{S}_n.
%\end{equation*}
%\item
%\end{enumerate}

\begin{remark}[\textbf{On degenerated and non-degenerated possible partitions}] \label{remark:onSizesOfBadGoodIndices}
One of the possible option resulting from
Conjecture~\ref{conjecture:OnTheBadChannels} is that $\mathcal{S}'_n
= \mathcal{S}_n$. In case where this degenerated partition is proved
to be correct, then it follows that the additional information
provided by the bits in $\mathbf{W}_{\mathcal{D}^{\text{c}}_n}$ do
not alter the known polarization of the mutual information
quantities $I_i$ in~\eqref{equation:mutualInformationQuantities}.
The non-degenerated partition of $\mathcal{S}_n$ offers (in the case
it is proven to be correct) a dichotomy of the indices in
$\mathcal{S}_n$. Accordingly, either the former polarization remains
or alternatively the knowledge of the bits in
$\mathbf{W}_{\mathcal{D}^{\text{c}}_n}$ completely changes the
orientation of the polarization. The size of $\mathcal{S}''_n \cup
\mathcal{A}'_n$ must satisfy
\begin{equation} \label{equation:boundingTheGoodOfBadSetSize}
|\mathcal{S}''_n \cup \mathcal{A}'_n| \stackrel{\text{(a)}} =
|\mathcal{A}'_n| + |\mathcal{A}''_n| \stackrel{\text{(b)}} \leq n
C(p).
\end{equation}
Equality~(a) in~\eqref{equation:boundingTheGoodOfBadSetSize} is
obvious as the sets $\mathcal{A}'_n$ and $\mathcal{S}_n$ are
disjoint. Violating the inequality~(b)
in~\eqref{equation:boundingTheGoodOfBadSetSize} results in violating
the coding theorem for a DMC as the input bits to the split channels
specified by the set $\mathcal{S}''_n \cup \mathcal{A}'_n$ can be
reliably decoded (This can be shown in a similar fashion as
in~\cite{ArikanPolarCodes}).
\end{remark}

\begin{remark}[\textbf{On a particular trivial case where
Conjecture~\ref{conjecture:OnTheBadChannels} is true}] There exists
an option where Conjecture\ref{conjecture:OnTheBadChannels} is
trivially proved as a particular application of
Theorem~\ref{theorem:PolarizationOfMutualInformation}. Specifically,
assume that for every index $i \in \mathcal{D}_n$, it follows that
\begin{equation*}
j < i \quad \forall j \in \mathcal{D}_n^{\text{c}}.
\end{equation*}
In that case, the degenerated partition in
Remark~\ref{remark:onSizesOfBadGoodIndices} follows as an immediate
particular case of
Theorem~\ref{theorem:PolarizationOfMutualInformation}.
\end{remark}

\subsection{A polar secrecy scheme}
\label{section:polarSecrecySchemeGeneralCase}

In this section, a polar secrecy scheme is provided assuming that
Conjecture~\ref{conjecture:OnTheBadChannels} is true. The same
notation and definitions of the coset code defined in
Section~\ref{section:polarCoding4DegradedWireTapChannel} are
assumed. The transmitted codeword $\mathbf{x}$ is defined
in~\eqref{equation:scureCodeword}. This definition is based on the
index sets $\mathcal{A}_n$ and $\mathcal{N}_n$. The secure
information bits are considered as if they are being transmitted
over the split channels whose indices are in $\mathcal{A}_n$. Over
the split channels whose indices are in $\mathcal{N}_n$, noisy bits
are attributed. The polar secrecy scheme is provided in
Section~\ref{section:TheProposedScheme} by a proper choice of the
sets $\mathcal{A}_n$ and $\mathcal{N}_n$. The degradation property
in Section~\ref{section:TheProposedScheme} assures that the indices
which correspond to split channels which polarize to `good channels'
for the eavesdropper, also polarize for `good channels' for the
legitimate user. This clearly does not necessarily follow for the
general not-degraded case.

For the general wiretap channel, indices that are `good' for the
eavesdropper may not be `good' for the legitimate user and
vice-versa. A binary-input symmetric wiretap channel is assumed. As
in the construction detailed in Part~I of the proof of
Theorem~\ref{theorem:secrecyAchevingDegradedCase}, the sets
$\tilde{\mathcal{A}}_n$ and $\tilde{\mathcal{N}}_n$ of `good
indices' are considered. The sets $\tilde{\mathcal{A}}_n$ and
$\tilde{\mathcal{N}}_n$ include the indices for which the
Bhattacharyya parameters of the corresponding split channels
approach zero as the block length approach infinity. Specifically,
fixing $r < C(P_{Y|X})$ and $r^* < C(P_{Z|X})$, the conditions
in~\eqref{equation:CardinalityPropertyEavesdropper}-\eqref{equation:BConstantPropertyLegitimate}
follow.

Define the index set $\mathcal{S}_n \triangleq \tilde{\mathcal{A}}_n
\setminus \tilde{\mathcal{N}}_n$ of indices which are `good' for
both the legitimate user and the eavesdropper. According to
Conjecture~\ref{conjecture:OnTheBadChannels}, the set
$\mathcal{S}_n$ can be partitioned into two index sets
$\mathcal{S}'_n$ and $\mathcal{S}''_n$, satisfying the polarization
properties in~
\eqref{equation:conditionOnSTag}-\eqref{equation:conditionOnSTagim}
where $\mathcal{A}_n$ is replaced by $\tilde{\mathcal{N}}_n$, and
$\mathcal{A}'_n$ is replaced by $\tilde{\mathcal{A}}_n \cap
\tilde{\mathcal{N}}_n$. Next, the set $\mathcal{N}_n$ is defined
according to
\begin{equation} \label{equation:NoisyBitIndices4GeneralCaseCondition2}
\mathcal{N}_n \triangleq \bigl(\tilde{\mathcal{A}}_n \cap
\tilde{\mathcal{N}}_n\bigr) \cup \mathcal{S}''_n
\end{equation}
and the set $\mathcal{A}_n$ is defined to be the remaining indices
in $\mathcal{S}_n$, that is
\begin{equation*}
\mathcal{A}_n \triangleq \mathcal{S}'_n.
\end{equation*}
As explained in Remark~\ref{remark:onSizesOfBadGoodIndices}, the
term $\frac{1}{n}|\mathcal{N}_n|$ can not exceed the capacity of the
eavesdropper marginal channel. Consequently, the size of
$\mathcal{S}'_n$ can be chosen such that
$\frac{1}{n}|\mathcal{S}'_n|$ is arbitrarily close to $C(P_{Y|X}) -
C(P_{Z|X})$.

%In the particular case where a degne The construction of the general
%caseprovided scheme depends on which part of
%Conjecture~\ref{conjecture:OnTheBadChannels} is true:
%\begin{enumerate}
%\item The case where the
%first part in Conjecture~\ref{conjecture:OnTheBadChannels} is true:
%The set $\mathcal{A}_n$ is chosen from the indices in
%$\tilde{\mathcal{A}}_n \setminus \tilde{\mathcal{N}}_n$ (it is
%favorable but not necessary to chose the indices of the split
%channels whose Bhattacharayya constants are minimal). Note that
%$\frac{1}{n}|\tilde{\mathcal{A}}_n|$ may be chosen arbitrarily close
%to $C(P_{Y|X})$ and the rate $\frac{1}{n}|\tilde{\mathcal{N}}_n|$
%can not be greater than $C(P_{Z|X})$ (otherwise this contradicts the
%converse of the channel coding theorem). Consequently, the
%cardinality of the set $\mathcal{A}_n$ can be chosen such that
%$\frac{1}{n}|\mathcal{A}_n|$ can be made arbitrarily close to
%$C(P_{Y|X}) - C(P_{Z|X})$. The set $\mathcal{N}_n$ is chosen to
%include all the indices that are `good' both to the legitimate user
%and the eavesdropper. That is,
%\begin{equation} \label{equation:NoisyBitIndices4GeneralCase}
%\mathcal{N}_n \triangleq \tilde{\mathcal{A}}_n \cap
%\tilde{\mathcal{N}}_n.
%\end{equation}
%
%\item The case where the second part of
%Conjecture~\ref{conjecture:OnTheBadChannels} is true:
%\end{enumerate}

Next, the same coset coding scheme defined
in~\eqref{equation:scureCodeword} is applied to the case at hand
(with the new construction of the sets $\mathcal{A}_n$ and
$\mathcal{N}_n$). As the information rate
$\frac{1}{n}|\mathcal{A}_n|$ of the considered scheme may be chosen
arbitrarily close to $C(P_{Y|X}) - C(P_{Z|X})$, the same coding rate
as in Theorem~\ref{theorem:secrecyAchevingDegradedCase} is obtained.
The decoding reliability at the legitimate user is clear and follows
the same proof as for the degraded case (note that all the noisy
bits in the considered scheme are `transmitted' over the split
channels that are `good' for the legitimate user). It is left to
establish that the equivocation rate can approach the information
rate of the considered scheme.

\subsection{Analysis of the equivocation rate}

As explained in
Section~\ref{section:polarCoding4DegradedWireTapChannel}, the bits
$\mathbf{b}_n$ corresponding to the indices in $\mathcal{B}_n$ are
predetermined and fixed. These bits are known both to the
eavesdropper and the legitimate user. For each blocklength $n$,
consider the ensemble of coset codes corresponding for all the
possible selection of fixed bits $\mathbf{b}_n$.
%As the considered wiretap channel is symmetric, the polarization
%properties\footnote{Both the regular polarization and the
%conjectured polarization properties are considered here. As for the
%regular polarization, the independence properties for symmetric
%channels is well proven in~\cite{ArikanPolarCodes}. As for
%independence properties for the conjectured polarization effect,
%such a result is currently a reasonable assumption.} of the split
%channels, for both the legitimate user and the eavesdropper are
%irrespective with the actual choice of the fixed bits
%$\mathbf{b}_n$.
An analysis of the equivocation rate where the coset code is chosen
in random is considered. Specifically, it is assumed that the actual
code is chosen from the ensemble by picking the bits in
$\mathbf{b}_n$ in random. The random selection of the bits in
$\mathbf{b}_n$ is carried independently and identically. Each bit is
picked at random with an equiprobable probability,
$\Pr(0)=\Pr(1)=\frac{1}{2}$. In addition, it is assumed that the
random selection of $\mathbf{b}_n$ is independent with the random
noisy bits in $\mathbf{b}^*_n$ and the secret message. It is
important to distinguish between the ransom selection of a code and
the noisy bits $\mathbf{b}^*$. The random selection of code is part
of our analysis, this selection (i.e., the bits in $\mathbf{b}_n$)
is known to both the legitimate and the eavesdropper. In contrast,
the random noisy bits $\mathbf{b}^*$ are immanent part of the
encoding procedure and they are unknown to both the legitimate user
and the receiver. The noisy bits $\mathbf{b}^*$ are picked randomly,
each independent with the others, and with a uniform probability.
The information bits are also assumed to be independent and
equiprobable.

The secrecy properties of the suggested scheme is considered in the
following proposition:

\begin{proposition}
\label{proposition:EquivocationRatePropertyGeneralCase} {\em
Consider the polar secrecy scheme in
Section~\ref{section:polarSecrecySchemeGeneralCase} whose
transmissions take place over a binary-input memoryless symmetric
wiretap channel. Then, there exists a bit vector $\mathbf{b}_n$ for
which the equivocation rate satisfy the secrecy condition
in~\eqref{equation:equivocationRatePropertyInTheorem}.}
\end{proposition}

\begin{proof}
Denote by $\mathbf{W}$ the random binary vector comprises the random
bits in $\mathbf{b}_n$, $\mathbf{b}^*_n$, and $\mathbf{u}$ in the
encoding procedure~\eqref{equation:scureCodeword}, and by
$\mathbf{Z}$ the random vector received at the eavesdropper.
According to the considered assumptions, all the bits in
$\mathbf{W}$ are independent and equiprobable. It follows using the
chain rule of mutual information that
\begin{align}
I(\mathbf{W}_{\mathcal{N}_n}, \mathbf{W}_{\mathcal{A}_n};
\mathbf{W}_{\mathcal{B}_n}, \mathbf{Z}) & =
I(\mathbf{W}_{\mathcal{A}_n}; \mathbf{W}_{\mathcal{B}_n},
\mathbf{Z}) + I(\mathbf{W}_{\mathcal{N}_n};
\mathbf{W}_{\mathcal{B}_n}, \mathbf{Z}\ | \
\mathbf{W}_{\mathcal{A}_n})
\nonumber \\
& = I(\mathbf{W}_{\mathcal{A}_n}; \mathbf{W}_{\mathcal{B}_n}) +
I(\mathbf{W}_{\mathcal{A}_n}; \mathbf{Z}\ | \
\mathbf{W}_{\mathcal{B}_n}) \nonumber \\
& + I(\mathbf{W}_{\mathcal{N}_n}; \mathbf{W}_{\mathcal{B}_n}\ | \
\mathbf{W}_{\mathcal{A}_n} ) + I(\mathbf{W}_{\mathcal{N}_n};
\mathbf{Z}\ | \ \mathbf{W}_{\mathcal{A}_n}, \mathbf{W}_{\mathcal{B}_n})\nonumber \\
\label{equation:stepInEquivocationRateAnalysisConj1} & =
H(\mathbf{W}_{\mathcal{A}_n}) - H(\mathbf{W}_{\mathcal{A}_n}|\
\mathbf{Z}, \mathbf{W}_{\mathcal{B}_n}) +
I(\mathbf{W}_{\mathcal{N}_n}; \mathbf{Z}\ | \
\mathbf{W}_{\mathcal{A}_n}, \mathbf{W}_{\mathcal{B}_n})
\end{align}
where the last equality follows since $\mathbf{W}_{\mathcal{A}_n}$,
$\mathbf{W}_{\mathcal{N}_n}$, and $\mathbf{W}_{\mathcal{B}_n}$ are
independent. As the set $\mathcal{N}_n$ comprises indices of split
channels which polarize to perfect channels, the bits in
$\mathbf{W}_{\mathcal{N}_n}$ can be reliably decoded at the
eavesdropper based on perfect knowledge of the remaining bits and
the received vector (this is shown in a similar fashion
to~\cite{ArikanPolarCodes}). Hence, the decoding error probability
$P_{\text{e}}(\mathbf{W}_{\mathcal{N}_n^{\text{c}}})$ of the bits in
$\mathbf{W}_{\mathcal{N}_n}$ based on the received vector and the
remaining bits $\mathbf{W}_{\mathcal{N}_n^{\text{c}}}$, can be made
arbitrarily low. As a consequence of Fano's inequality it follows
that
\begin{align}
|\mathcal{N}_n| & \geq I(\mathbf{W}_{\mathcal{N}_n}; \mathbf{Z}\ | \
\mathbf{W}_{\mathcal{A}_n}, \mathbf{W}_{\mathcal{B}_n}) \nonumber\\
& = H(\mathbf{W}_{\mathcal{N}_n}|\ \mathbf{W}_{\mathcal{A}_n},
\mathbf{W}_{\mathcal{B}_n}) - H(\mathbf{W}_{\mathcal{N}_n}|\
\mathbf{Z}, \mathbf{W}_{\mathcal{A}_n}, \mathbf{W}_{\mathcal{B}_n})
\nonumber\\
\label{equation:step2InEquivocationRateAnalysisConj1} & >
H(\mathbf{W}_{\mathcal{N}_n}) - h_2  \bigl(
P_{\text{e}}(\mathbf{W}_{\mathcal{N}_n^{\text{c}}})\bigr) -
|\mathcal{N}_n| P_{\text{e}}(\mathbf{W}_{\mathcal{N}_n^{\text{c}}})
\end{align}
where $h_2$ is the binary entropy function. For a sufficiently large
block length $n$, the expected decoding error probability approaches
zero. Consequently, $\frac{1}{n} I(\mathbf{W}_{\mathcal{N}_n};
\mathbf{Z}\ | \ \mathbf{W}_{\mathcal{A}_n},
\mathbf{W}_{\mathcal{B}_n})$ can be made arbitrarily close to
$\frac{1}{n}|\mathcal{N}_n|$. It follows
from~\eqref{equation:stepInEquivocationRateAnalysisConj1}
and~\eqref{equation:step2InEquivocationRateAnalysisConj1} that
\begin{equation} \label{equation:step3InEquivocationRateAnalysisConj1}
\frac{1}{n} H(\mathbf{W}_{\mathcal{A}_n}|\ \mathbf{Z},
\mathbf{W}_{\mathcal{B}_n}) \geq \frac{|\mathcal{A}_n|}{n}  +
\frac{|\mathcal{N}_n|}{n}  - \epsilon_n - \frac{1}{n}
I(\mathbf{W}_{\mathcal{N}_n}, \mathbf{W}_{\mathcal{A}_n};
\mathbf{W}_{\mathcal{B}_n}, \mathbf{Z})
\end{equation}
where $\epsilon_n \geq 0$ and approaches zero as $n$ grows.

Based on Conjecture~\ref{conjecture:OnTheBadChannels}, the mutual
information $\frac{1}{n} I(\mathbf{W}_{\mathcal{N}_n},
\mathbf{W}_{\mathcal{A}_n}; \mathbf{W}_{\mathcal{B}_n}, \mathbf{Z})$
can be shown to be arbitrarily close to $\frac{1}{n}
|\mathcal{N}_n|$. Using the chain rule of mutual information it
follows that
\begin{align}
I(\mathbf{W}_{\mathcal{N}_n}, \mathbf{W}_{\mathcal{A}_n};
\mathbf{W}_{\mathcal{B}_n}, \mathbf{Z}) & = \sum_{i \in
\mathcal{N}_n} I(\mathbf{W}_i; \mathbf{W}_{\mathcal{B}_n},
\mathbf{Z}\ | \ \mathbf{W}_{\mathcal{N}^{(i)}_n},
\mathbf{W}_{\mathcal{A}^{(i)}_n}) \nonumber\\
& + \sum_{i \in \mathcal{A}_n} I(\mathbf{W}_i;
\mathbf{W}_{\mathcal{B}_n}, \mathbf{Z}\ | \
\mathbf{W}_{\mathcal{N}^{(i)}_n}, \mathbf{W}_{\mathcal{A}^{(i)}_n}).
\nonumber \\
& = \sum_{i \in \mathcal{N}_n} I(\mathbf{W}_i;
\mathbf{W}_{\mathcal{N}^{(i)}_n}, \mathbf{W}_{\mathcal{A}^{(i)}_n},
\mathbf{W}_{\mathcal{B}_n}, \mathbf{Z}) \nonumber\\
\label{equation:chainRuleStep} & + \sum_{i \in \mathcal{A}_n}
I(\mathbf{W}_i; \mathbf{W}_{\mathcal{N}^{(i)}_n},
\mathbf{W}_{\mathcal{A}^{(i)}_n}, \mathbf{W}_{\mathcal{B}_n},
\mathbf{Z}).
\end{align}
where the last equality follows as all the bits in $\mathbf{W}$ are
independent. For every index $i\in \mathcal{N}_n$, it follows from
Lemma~\ref{lemma:OnModifiedPolarizationOfGoodBits} and
Conjecture~\ref{conjecture:OnTheBadChannels} that
\begin{equation} \label{eqaution:noisyBitsPolarizationBoundOption1}
I(W_i; \mathbf{W}_{\mathcal{N}_n^{(i)}},
\mathbf{W}_{\mathcal{A}_n^{(i)}}, \mathbf{W}_{\mathcal{B}_n},
\mathbf{Z}) > 1-\delta.
\end{equation}
In addition, for all the indices $i \in \mathcal{A}_n$ it also
follows from Conjecture~\ref{conjecture:OnTheBadChannels} that
\begin{equation} \label{equation:badBitsPolarizationBoundOption1}
I(W_i; \mathbf{W}_{\mathcal{N}_n^{(i)}},
\mathbf{W}_{\mathcal{A}_n^{(i)}}, \mathbf{W}_{\mathcal{B}_n},
\mathbf{Z}) < \delta.
\end{equation}
From
\eqref{equation:chainRuleStep},~\eqref{eqaution:noisyBitsPolarizationBoundOption1}
and~\eqref{equation:badBitsPolarizationBoundOption1} it follows that
\begin{align}
\nonumber \frac{1}{n} I(\mathbf{W}_{\mathcal{N}_n},
\mathbf{W}_{\mathcal{A}_n}; \mathbf{W}_{\mathcal{B}_n}, \mathbf{Z})
& \leq \frac{|\mathcal{N}_n|}{n} + \frac{\delta |\mathcal{A}_n|}{n}\\
\label{equation:almostLastStepOfGeneralCase} & \leq
\frac{|\mathcal{N}_n|}{n} + \delta.
\end{align}
Hence, based
on~\eqref{equation:step3InEquivocationRateAnalysisConj1}
and~\eqref{equation:almostLastStepOfGeneralCase} we end up with
\begin{equation*}
\frac{1}{n} H(\mathbf{W}_{\mathcal{A}_n}|\ \mathbf{Z},
\mathbf{W}_{\mathcal{B}_n}) \geq \frac{1}{n} |\mathcal{A}_n|
 - \epsilon_n
-\delta.
\end{equation*}
As $\delta$ can be fixed arbitrarily small, and $\epsilon_n$
approaches zero, the equivocation rate can be made arbitrarily close
to $\frac{1}{n} |\mathcal{A}_n|$ which assures the secrecy property
of the provided scheme.
\end{proof}

\section{Summery and Conclusions} \label{section:generalizations}

A polar secrecy scheme is provided in this paper for the two-user,
memoryless, symmetric and degraded wire-tap channel.
The provided polar codes are shown to achieve the entire rate-equivocation region.
Our polar coding scheme is based on the
channel polarization method originally introduced by Arikan~\cite{ArikanPolarCodes}
for single-user setting. For the particular case of binary erasure channel, the secrecy capacity is shown to achieve the secrecy capacity under the strong notion of secrecy.

Proving (disproving, or finding a counter example) Conjecture~\ref{conjecture:OnTheBadChannels} is the main
interest in the continuation of the research discussed in this
paper. The following generalizations are of additional possible
interest:
\begin{enumerate}
\item
Non-binary settings: In light of the recent results by Sasoglu et
al.~\cite{NonBinaryPolar}, a generalization to the non-binary
setting may be a straight forward generalization.
\item Secrecy polar schemes for non-symmetric wiretap channels, based on the non-binary polarization
provided in~\cite{NonBinaryPolar}.
\item Polar coding for a broadcast channel with confidential messages. The particular case of
degraded message sets over a degraded channel is first considered.
\item Strong secrecy properties: As noted, the provided scheme is
shown to provide weak secrecy. It is of great interest to find out
if this scheme can also provide strong secrecy.
\item Generalized polar secrecy-schemes based on the ideas
in~\cite{Arikan2D}, \cite{Satish1}-\cite{KoradaPHD}.
\item Combing the polar scheme with the MAC approach for the
wiretap channel (see, e.g., \cite{MACApproach}).
\end{enumerate}

\subsection*{Acknowledgment}
The authors are grateful to Prof. Emre Telatar for reviewing an early version of this paper in October~2009.

\end{document}